\documentclass[journal, 12pt, draftclsnofoot, onecolumn]{IEEEtran}
\IEEEoverridecommandlockouts

\usepackage{amsfonts,amssymb}
\usepackage{ifpdf}
\usepackage{cite}
\usepackage{stfloats}
\usepackage{bm}
\usepackage{color,xcolor}
\usepackage{subfigure}
\ifCLASSINFOpdf
   \usepackage[pdftex]{graphicx}
   \graphicspath{{../pdf/}{../jpeg/}}
   \DeclareGraphicsExtensions{.pdf,.jpeg,.png}
\else
   \usepackage[dvips]{graphicx}

   \graphicspath{{../eps/}}

   \DeclareGraphicsExtensions{.eps}
\fi

\usepackage[cmex10]{amsmath}

\usepackage{algorithm}
\usepackage{algorithmic}
\ifCLASSOPTIONcompsoc
\else
  \usepackage[caption=false,font=footnotesize]{subfig}
\fi

\usepackage{makecell}
\begin{document}
\title{Joint  Active and Passive  Beamforming Design for IRS-Aided   Radar-Communication}
\author{Meng~Hua,
	Qingqing~Wu,~\IEEEmembership{Senior Member,~IEEE,}
	Chong He,~\IEEEmembership{Member,~IEEE,} 
	Shaodan~Ma,~\IEEEmembership{Senior Member,~IEEE,}
and Wen~Chen,~\IEEEmembership{Senior Member,~IEEE}
\thanks{M. Hua, Q. Wu, and  S. Ma are with the State Key Laboratory of Internet of Things for Smart City, University of Macau, Macao 999078, China (email: menghua@um.edu.mo; qingqingwu@um.edu.mo; shaodanma@um.edu.mo). }
\thanks{ C. He and  W. Chen are with the Department of Electronic Engineering, Shanghai Institute of Advanced Communications and Data Sciences,
	Shanghai Jiao Tong University, Minhang 200240, China (e-mail: hechong@sjtu.edu.cn; wenchen@sjtu.edu.cn).}
}
\maketitle
\vspace{-1cm}
\begin{abstract}
In this paper, we study an intelligent reflecting surface (IRS)-aided radar-communication (Radcom) system, where the IRS is leveraged to help Radcom base station (BS)   transmit the joint of  communication signals and radar signals for serving  communication users  and tracking targets simultaneously. The objective of this paper is  to  minimize the total transmit power   at the Radcom BS by jointly optimizing the active beamformers, including communication beamformers and radar beamformers,  at the Radcom BS and  the phase shifts  at the IRS,  subject to the minimum  signal-to-interference-plus-noise ratio (SINR) required by communication users, the minimum SINR required by the radar,  and the cross-correlation pattern design. In particular, we consider  two  cases, namely, case I and case II, based on  the presence or absence of  the radar cross-correlation  design   and   the  interference introduced by the IRS   on  the Radcom BS. For   case I where the cross-correlation design and the interference are not considered, we  prove that the dedicated radar signals are not  needed, which significantly  reduces implementation
complexity  and simplifies algorithm design. Then, a penalty-based algorithm is proposed to solve the resulting   non-convex optimization problem. Whereas for  case II considering   the cross-correlation design and the interference,  we unveil that the dedicated radar signals  are  needed in general to enhance the system performance. Since the resulting   optimization problem is more challenging to solve as compared with the case I, the semidefinite relaxation (SDR)  based alternating optimization (AO) algorithm  is proposed. Particularly, instead of relying on the Gaussian randomization  technique to obtain an approximate solution by reconstructing rank-one solution,     the tightness  is achieved by our proposed reconstruction strategy. 
 Simulation results demonstrate the effectiveness of  proposed algorithms and also show the superiority of the proposed  scheme  over various benchmark schemes. 
\end{abstract}
~\\
\section{Introduction}
The rapid increase of mobile data  and  Internet of Things (IoT) devices are creating unprecedented challenges for wireless service providers to  provide high data rate and ultra-reliable low latency communication due to the limited  frequency  spectrum from  $700~{\rm MHz}$ and $2.6~{\rm GHz}$ in the existing communication networks \cite{Rappaport2013Millimeter}. In contrast,  the radar system  has fruitful  spectrum resource and typically operate ranging  from $0.3{\text-}100~{\rm GHz}$, such as S band $(2{\text-}4~{\rm GHz})$, C band  $(4{\text-}8~{\rm GHz})$, and X band  $(8{\text-}12.5~{\rm GHz})$,  depending on specific application requirements \cite{parker2017digital}. The coexistence (or spectrum sharing) design  between
the radar system and the  wireless communication system is attracting great attention, which allows the communication system to  use  the spectrum resource of the radar system \cite{zheng2019Radar}. To  mitigate the interference   between the above two  systems, several promising  approaches are proposed, such as the opportunistic spectrum sharing approach \cite{Saruthirathanaworakun2012Opportunistic}, the  null-space projection based approach  \cite{mu2018liu,Sodagari2012projection}, and the joint design of  radar waveform and communication  beamforming  \cite{li2016optimum,li2016mimoconference}.
However,  such  separated deployment, i.e., the  radar transceiver and the communication transmitter are geographically separated,  requires additional information, such   as the channel state information (CSI), radar probing waveforms, and communication modulation
format, etc., to exchange   to coordinate the simultaneous radar and communication transmissions, which significantly increases  the  complexity for  hardware implementation in practice.

The radar-communication (Radcom) system (also known as the dual-function radar-communication system \cite{Hassanien2016Signaling}), which integrates the  radar and communication functions into a single hardware platform and  
 is  regarded as a promising solution to simplify the system design \cite{Sturm2011waveform}. The RadCom system is able to  simultaneously perform both radar and communication functionalities using the same signals  transmitted from a fully-shared transmitter, which does not require to exchange information  and naturally achieves full cooperation. 
 In the early stage, the information is   embedded into the radar pulses so that the communication transmission can be readily realized by using the already  fabricated radar platforms. For example, the communication symbols   embedding into radar pulses and/or sidelobe can   be realized by controlling the radar pulse's amplitude, phase shift, and even index modulation \cite{Hassanien2016dual, hassanien2016phase,csahin2021index}. However, such  approaches result in a low data rate for transmission  since the    transmission rate  is  fundamentally constrained by the radar  pulse repetition frequency. 
Another important paradigm of research in existing works on Radcom is the transmit beamforming design \cite{mu2018liu,liu2018towards,liu2021dual}. Compared to  the  information embedding approaches, the  transmit beamforming design   potentially
supports high  data rate  and guarantees radar performance
by   synthesizing a  joint waveform   that is shared by both radar and communications. The  seminal work in \cite{mu2018liu} analyzed the synthesized waveform performance in the   shared deployment system, and showed that the shared deployment significantly outperforms than the separated deployment in terms of the trade-off between the radar beampattern synthesis  and the quality of communication. Instead of using the synthesized waveforms, the authors in \cite{joint2018liu} 
proposed to   transmit the combination signals with   communication  beamformers and 
 radar waveforms at the Radcom base station (BS), where the communication  beamformers are  used for serving communication users and radar waveforms are used for radar sensing, which     provides more degrees of freedom for system design. The results in \cite{joint2018liu} showed that the communication-only beamformer design is inferior to the joint design of communication beamformer and radar waveform in terms of beampattern synthesis, especially when the number of communication users is less than the number of targets. The authors in  \cite{hua2021optimal} further answered  whether radar waveforms are needed under different design criteria, channel conditions,  and receiver types.
 However,  the above works  focused on either the beamforming/waveform design or  encoding design, the limited degrees of freedom such as the uncontrollable of  electromagnetic waves  propagation   still  confine the system performance. 

Recently, intelligent reflecting surfaces (IRSs) attract  great attention both from industry and academia \cite{WU2020towards,wu2020intelligentarxiv,liu2021Reconfigurable}. The IRS is composed of large numbers of  passive and low-cost reflecting elements, each of which is able to 
 independently adjust  phase shift and/or amplitude on the impinging electromagnetic signals, so that  it is able 
 to controllably  change  the electromagnetic waves  propagation  towards any directions of interest. The seminal work in \cite{wu2019intelligentxx} unveiled the  fundamental scaling law of the IRS by showing that    the received signal-to-noise (SNR) is  quadratically  increasing with the number of IRS reflecting elements. Based on this appealing result, the IRS has been exploited  for different applications such as wireless information transmission \cite{pan2020multicell,hua2020intelligent,zhou2020intelligent}, wireless-powered communication network \cite{wu2021irs,mengjoint2021,chen2021irsaided}, unmanned aerial vehicle communication  \cite{hui2019Reflections,hua2021UAVSymbiotic,sixian2020robust}, and non-orthogonal multiple access \cite{mu2020Exploiting,9365004,9380234}, etc. 
To unleash the full  potential of the  Radcom system for both communication and radar sensing, the integration of  IRS in   the Radcom system  is also ongoing. A handful of works, see e.g., \cite{buzzi2021foundations,Lu2021Target,vaca2021radio, buzzi2021Radar}, studied the fundamental problem of target detection with the help of IRS in the radar-only system.
 Some further works, see, e.g.,   \cite{9364358,song2021joint,9416177,liu2021joint}, focused on the Radcom   system by exploiting IRS to enhance the sensing performance while satisfying the quality-of-service (QoS) of users. The authors in \cite{9364358} and \cite{song2021joint} studied    one communication user scenario. To be specific, work \cite{9364358} studied one target sensing and 
aimed at maximizing the radar SNR while guaranteeing  the QoS of the user by the joint optimization of  IRS   phase shifts and transmit covariance matrix.  In \cite{song2021joint}, the authors  studied the scenario where the BS and multiple sensing targets are blocked and constructed a virtual  line-of-sight (LoS) link between the IRS and   targets for target sensing.
The authors in \cite{9416177} and \cite{liu2021joint} further studied  the joint design of the active beamforming and the passive
beamforming for the multi-user scenario in RIS-assisted Radcom system. The  minimization of   multi-user interference  under the predefined beampattern constraint was studied in \cite{9416177}. 
 The authors in \cite{liu2021joint} studied  the scenario that the target sensing is corrupted by  multiple  clutters with the  signal-dependent interference  and aimed to maximize  the radar SINR. However, the above works ignore the impact of   interference introduced by the IRS and  radar cross-correlation on the system. 
 In addition,   regarding the   transceiver design for the joint waveform design or the  single waveform design is also  not answered and studied.
 \begin{figure}[!t]
 	\centerline{\includegraphics[width=2.5in]{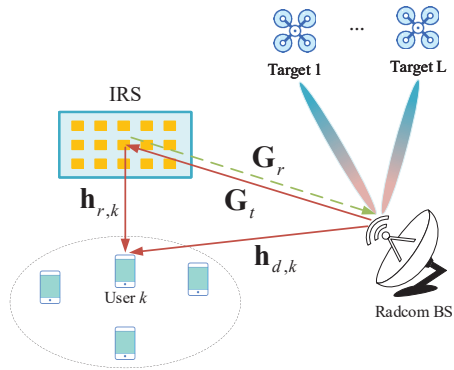}}
 	\caption{An IRS-aided Radcom system.} \vspace{-0.7cm}\label{fig1}
 \end{figure}

To address the above challenges, this paper studies an IRS-aided Radcom system in multi-user and multi-target scenarios  as shown in Fig.~\ref{fig1}.
Based on  the presence or absence of  the radar cross-correlation  design   and   the  interference introduced by the IRS   on  the Radcom BS, two cases, namely, case I and case II, are studied.
In particular, we answer the fundamental question: whether the dedicated radar signals are needed  for these two cases?
The main contributions of this paper are summarized as follows.
\begin{itemize}
	\item We study an IRS-aided  Radcom  system, where the IRS is leveraged to help the Radcom BS   transmit the joint of  communication signals and radar signals for serving  communication users  and tracking targets.  Our objective is to  minimize the total transmit power   at the Radcom BS by jointly optimizing the active beamformers at the Radcom BS and  the phase shifts at the IRS,  subject to the minimum  signal-to-interference-plus-noise ratio (SINR) required by communication users, the  minimum SINR required by  the radar, and the cross-correlation pattern design. In particular, we consider  two  cases, i.e., case I and case II, based on  the presence or absence of  the radar cross-correlation  design   and   the  interference introduced by the IRS   on  the Radcom BS, which results in  two different optimization problems. 
\item For  case I  where the interference introduced by   the IRS is   perfectly    canceled and     the cross-correlation design is ignored. Since the resulting optimization problem is non-convex, there
are no standard convex methods to solve it optimally. To solve the problem, 
we  first rigorously prove that the dedicated radar signals are not  needed in this case, which significantly  reduces implementation complexity  and simplifies the following  algorithm design.  Then,  we propose a novel penalty-based algorithm, which includes a two-layer iteration, i.e., an inner layer iteration and an outer layer iteration. The inner layer solves the penalized optimization problem,
while the outer layer updates the penalty coefficient over iterations to guarantee convergence.
In particular,  the solution to  each subproblem in the  inner layer is solved by either a closed-form expression or a semi-closed-form expression.
\item For  case II  where both the cross-correlation pattern design  and the interference introduced  by the IRS  are considered. Since the resulting optimization problem is more challenging to solve than  case I, the proposed penalty-based algorithm in case I cannot be applicable to this case. To solve this difficulty, a    semidefinite relaxation (SDR)-based alternating optimization (AO) is proposed. In particular, instead of relying on the Gaussian randomization  technique to obtain an approximation solution,       the tightness  is achieved by our proposed reconstruction strategy. 
In addition, we unveil that the dedicated radar signals  are  needed in general for this case to enhance the system performance.
\item Our simulation results demonstrate that the IRS is beneficial for reducing the transmit power required by the Radcom BS in case I. In addition, it is shown that in case II, as the IRS is deployed far from the Radcom BS, the IRS is helpful for reducing transmit power, while as the IRS is deployed in the vicinity of the Radcom BS, the IRS may  even deteriorate the system performance due to the interference.  Furthermore, the results also show that adopting  dedicated radar signals  at the Radcom BS can significantly reduce the system outage probability as compared to the case without adopting the dedicated radar signals in case II.
\end{itemize} 
The rest of this paper is organized as follows. Section II introduces the system model and problem formulation  for the considered  IRS-aided Radcom system. In Section III,  a  penalty-based algorithm is proposed to solve   case I. In Section IV, an SDR-based AO algorithm is proposed to solve case II. 
 Numerical results are provided in Section V and  the paper is concluded in Section VI.

\emph{Notations}: Boldface upper-case and lower-case  letter denote matrix and   vector, respectively.  ${\mathbb C}^ {d_1\times d_2}$ stands for the set of  complex $d_1\times d_2$  matrices. For a   scalar value  $ x$, ${\left| { x} \right|}$ represents the  Euclidean norm of $ x$.
 For a  vector $\bf x$, ${\bf x}^*$ and  ${\bf x}^H$  stand for  its conjugate and  conjugate transpose, respectively,  and ${\rm diag}(\bf x) $ denotes a diagonal matrix whose main diagonal elements are extracted from vector $\bf x$.
 For a  matrix $\bf X$,  ${\rm{Tr}}\left( {\bf{X}} \right)$ and ${\rm{rank}}\left( {\bf{X}} \right)$   stand for  its trace and rank, respectively,  while ${\bf{X}} \succeq {\bf{0}}$ indicates that matrix $\bf X$ is positive semi-definite.
A circularly symmetric complex Gaussian random vector $\bf x$ with mean $ \bm \mu$ and covariance matrix  ${\bf{\Sigma }}$ is denoted by ${\bf{x}} \sim {\cal CN}\left( {\bm \mu ,{\bf{\Sigma }}} \right)$. ${\mathbb E}(\cdot)$ denotes the expectation operation. ${\cal O}\left(  \cdot  \right)$ is the big-O computational complexity notation.

\section{System Model and Problem Formulation}
\subsection{System Model}
Consider an IRS-aided Radcom system consisting of a Radcom BS, $K$ single-antenna users with the set denoted by  ${\cal K}=\{1,\dots,K\}$, $L$ radar targets with the set denoted by ${\cal L}=\{1,\dots,L\}$, and an IRS with $M$ reflecting elements  and with the set denoted by ${\cal M}=\{1,\dots,M\}$, as shown in Fig.~\ref{fig1}.   
The  Radcom BS is equipped with $N_t+N_r$ antennas, of which   $N_t$  transmit antennas are used for serving  communication users and tracking  radar targets at the same time, while   $N_r$ receive  antennas  are dedicated to receiving the echo signals reflected by radar targets. 
 \subsubsection{Transmit Waveform Design}
 The transmitted signal  by the Radcom BS is given by
 \begin{align}
{\bf{s}} = {{\bf{W}}_c}{{\bf{x}}_c} + {{\bf{W}}_r}{{\bf{x}}_r},
 \end{align}
where ${{\bf{x}}_c} \in {{\mathbb C}^{{K} \times 1}}$ denotes the transmit signals intended for  communication users   satisfying ${{\bf{x}}_c} \sim {\cal CN}\left( {{\bf{0}},{{\bf{I}}_K}} \right)$ and ${{\bf{W}}_c} \in {{\mathbb C}^{{N_t} \times {K}}}$ represents  the corresponding communication beamformer. Similarly, ${{\bf{x}}_r} \in {{\mathbb C}^{{N_t} \times 1}}$ denotes $N_t$ individual radar signals satisfying ${\mathbb E}\left\{ {{{\bf{x}}_r}} \right\} = {\bf{0}}_{N_t\times 1}$ and ${\mathbb E}\left\{ {{{\bf{x}}_r}{\bf{x}}_r^H} \right\} = {{\bf{I}}_{{N_t}}}$, and  ${{\bf{W}}_r} \in {{\mathbb C}^{{N_t} \times {N_t}}}$ represents the    radar beamformer. In addition, we assume that the  communication and radar signals are statistically independent  and uncorrelated, i.e., ${\mathbb E}\left\{ {{{\bf{x}}_c}{\bf{x}}_r^H} \right\} = {{\bf{0}}_{K \times {N_t}}}$ \cite{joint2018liu}.


\subsubsection{Communication Model}
 We consider a quasi-static flat-fading channel  in which the CSI remains unchanged in a channel coherence  block, but may change in the subsequent blocks.  We assume that the perfect  CSI of  all  involved  channels for communication is available at the Radcom BS via sending    pilot signals by  users \cite{wu2020intelligentarxiv}. Without loss of generality, in the downlink transmission, 
 denote by  ${\bf{G}}_t \in {{\mathbb C}^{M \times N_t}}$, ${\bf{h}}_{r,k}^H \in {{\mathbb C}^{1 \times M}}$, and ${{\bf h}^H_{d,k}} \in {{\mathbb C}^{1 \times N_t}}$ the complex equivalent baseband   channel between the Radcom BS and  the  IRS, between  the IRS  and the $k$th user, and  between  the  Radcom BS and the $k$th user,  $k\in{\cal K}$, respectively.  In addition, denote by  ${\bf{G}}_r \in {{\mathbb C}^{N_r \times M}}$ the complex equivalent baseband   channel between the Radcom BS and  the  IRS in the uplink transmission.
The received signal  by user $k$  in the downlink  is given by
\begin{align}
{y_k} = \left( {{\bf{h}}_{d,k}^H + {\bf{h}}_{r,k}^H{\bf{\Theta }}{{\bf{G}}_t}} \right){\bf{s}} + {n_k},
\end{align}
where  ${\bf{\Theta }} = {\rm{diag}}\left( {{v_1}, \cdots ,{v_M}} \right)$ represents  the  IRS reflection coefficient matrix and   $v_{m}$ denotes
the  phase shift  corresponding to the $m$th IRS reflecting element with $\left| {{v_m}} \right| = 1$, and $n_{k}\sim {\cal CN} \left( {0,\sigma ^2} \right)$ stands for the additive white Gaussian noise at user $k$. Define ${{\bf{W}}_c} = \left[ {{{\bf{w}}_{c,1}}, \ldots ,{{\bf{w}}_{c,K}}} \right]$, where ${{\bf{w}}_{c,k}} \in {{\mathbb C}^{{N_t} \times 1}}$ denotes the $k$th column vector of ${{\bf{W}}_c}, k\in {\cal K}$. Similarly, define ${{\bf{W}}_r} = \left[ {{{\bf{w}}_{r,1}}, \ldots ,{{\bf{w}}_{r,{N_t}}}} \right]$, where ${{\bf{w}}_{r,i}} \in {{\mathbb C}^{{N_t} \times 1}}$ denotes the  $i$th column vector of ${{\bf{W}}_r}$, $i\in {\cal N}_t=\{1,\dots,N_t\}$.
As such, the received SINR by user $k$ is given by 
\begin{align}
{\rm{SIN}}{{\rm{R}}_k} = \frac{{{{\left| {{\mathbf{h}}_k^H{{\mathbf{w}}_{c,k}}} \right|}^2}}}{{\sum\limits_{i \ne k}^K {{{\left| {{\mathbf{h}}_k^H{{\mathbf{w}}_{c,i}}} \right|}^2} + \sum\limits_{j = 1}^{{N_t}} {{{\left| {{\mathbf{h}}_k^H{{\mathbf{w}}_{r,j}}} \right|}^2}}  + {\sigma ^2}} }}, k \in {\cal K},
\end{align}
where ${\bf{h}}_k^H{\rm{ = }}{\bf{h}}_{d,k}^H + {\bf{h}}_{r,k}^H{\bf{\Theta G}}_t$. 

\subsubsection{Radar Model} 
Under the assumption that the propagation is nondispersive for tracking targets, the signal at the $l$th target location with angle $\theta_l$, $l\in{\cal L}$, can be described as $	{\bf{a}}_t^H\left( {{\theta _l}} \right){\bf{s}}$, where ${\bf{a}}_t^H\left( \theta  \right) \in {{\mathbb C}^{1 \times {N_t}}} = \left[ {1,{e^{ - j2\pi d\sin \left( \theta  \right)/\lambda }}, \ldots ,{e^{ - j2\pi d\left( {{N_t} - 1} \right)\sin \left( \theta  \right)/\lambda }}} \right]$ stands for the transmit steering vector at direction $\theta$ with $d$ denoting the antenna spacing and $\lambda$  denoting the carrier wavelength.
The received echo signals at the Radcom BS comes from four aspects, i.e., radar-target-radar channel, radar-IRS-radar channel, radar-IRS-target-radar channel, and radar-target-IRS-radar channel. However,  \cite{buzzi2021foundations}     showed by  both theoretical analysis and numerical simulations
 that the  signals   go  through  the  radar-target-IRS-radar channel and radar-IRS-target-radar channel are highly  attenuated due to  three-hop transmissions, which  has little impact on the system performance improvement.
Thus, we only need to consider the former two types of echo signals and  the received echo signals at the Radcom BS  is  given by
\begin{align}
{{\bf{y}}_r} = \underbrace {\sum\limits_{l = 1}^L {{\beta _l}{{\bf{a}}_r}\left( {{\theta _l}} \right){\bf{a}}_t^H\left( {{\theta _l}} \right)} {\bf{s}}}_{{\rm{reflected{\kern 1pt} {\kern 1pt}by{\kern 1pt} {\kern 1pt}targets}}} + \underbrace {{{\bf{G}}_r}{\bf{\Theta }}{{\bf{G}}_t}{\bf{s}}}_{{\rm{reflected{\kern 1pt} {\kern 1pt}by{\kern 1pt} {\kern 1pt}the{\kern 1pt} {\kern 1pt}IRS}}}{\rm{ + }}{{\bf n}_r}, \label{echo_signals}
\end{align}
where  $\beta _l$    represents the  reflection coefficient of the $l$th target,  which is  proportional
to the radar-cross section (RCS) of  the $l$th target\footnote{As stated in \cite{fishler2006spatial},  the  target is in general composed of an infinite number	of random, isotropic and independent scatterers over the area of interest, and the complex gain of the scatterer can be modeled as    a zero-mean and white complex random variable.  Together with the fact that incident angles between different targets are randomly distributed,  the amplitudes of different targets  can thus be assumed to be  independently distributed, i.e.,  ${\beta _l} \sim {\cal CN}\left( {0,\sigma _\beta ^2} \right)$ \cite{cui2014MIMO,cheng2018MIMO}.}  and  ${{\bf{n}}_r} \sim {\cal CN}\left( {{\bf{0}},{{  \sigma} ^2}{{\bf{I}}_{{N_r}}}} \right)$ stands for the additive white Gaussian noise at the Radcom BS. In addition, similar to ${\bf{a}}_t^H\left( \theta  \right)$,  ${{\bf{a}}_r}\left( \theta  \right) \in {{\mathbb C}^{{N_r} \times 1}}$ denotes the receive steering vector  at direction $\theta$. Here, two key points  need to be highlighted. First,  the communication signals, i.e., ${{\bf{W}}_c}{{\bf{x}}_c}$,  are not interference   for target tracking in the Radcom system (in other words, the communication signals are not  only used for downlink communication but also  used for target tracking) since the communication signals are known at the Radcom BS.  Second, since the reflected signals by IRS in \eqref{echo_signals} do not contain any information about the targets’ information, the radar SINR can be expressed as  \cite{li2016mimoconference,zheng2018joint}\footnote{ As stated in \cite{probing2007Stoica},  maximizing the radar SINR of the received signals is   a more justifiable goal than maximizing the total spatial power at a number of  given target locations. Thus, we consider the radar SINR as the design metric in this paper. In addition, we assume that targets' locations $\theta_l,l \in {\cal L}$ and amplitudes of channels ${\beta _l}, l\in {\cal L}$ are known at the Radcom BS for radar tracking by applying the effective estimation techniques such as   generalized likelihood ratio
	test (GLRT) and  Capon methods\cite{target2008xu,liu2006lu,liu2021dual}.} 
\begin{align}
{\rm{SIN}}{{\rm{R}}_r} = {\rm{tr}}\left( {{\bf{AR}}{{\bf{A}}^H}{{\left( {{\bf{BR}}{{\bf{B}}^H} + {\sigma ^2}{{\bf{I}}_{{N_r}}}} \right)}^{ - 1}}} \right), \label{radarsinr}
\end{align}
where ${\bf{R}} = \sum\limits_{k = 1}^K {{{\bf{w}}_{c,k}}{\bf{w}}_{c,k}^H + } \sum\limits_{i = 1}^{{N_t}} {{{\bf{w}}_{r,i}}{\bf{w}}_{r,i}^H} $,  ${\mathbf{A}}{\text{ = }}\sum\limits_{l = 1}^L {{\beta _l}{{\mathbf{a}}_r}\left( {{\theta _l}} \right){\mathbf{a}}_t^H\left( {{\theta _l}} \right)} $, and ${\bf{B}} = {{\bf{G}}_r}{\bf{\Theta G}}_t$. 
\subsection{Problem Formulation}
The objective of this paper is to minimize the total transmit power   at the Radcom BS by jointly optimizing the active beamformers at the Radcom BS and  the phase shifts at the IRS,  subject to the minimum SINR required by communication users, the minimum SINR required by radar, and cross-correlation pattern design.
 Accordingly, the optimization problem is formulated as
 \begin{subequations} \label{P1}
	\begin{align}
		&\mathop {\min }\limits_{{\{v_m\}},{\left\{ {{{\bf{w}}_{c,i}}} \right\},\left\{ {{{\bf{w}}_{r,i}}} \right\}}} \sum\limits_{k = 1}^K {{{\left\| {{{\bf{w}}_{c,k}}} \right\|}^2} + } \sum\limits_{i = 1}^{{N_t}} {{{\left\| {{{\bf{w}}_{r,i}}} \right\|}^2}} \\
		& {\rm s.t.}~\frac{{{{\left| {{\bf{h}}_k^H{{\bf{w}}_{c,k}}} \right|}^2}}}{{\sum\limits_{i \ne k}^K {{{\left| {{\bf{h}}_k^H{{\bf{w}}_{c,i}}} \right|}^2} + \sum\limits_{i = 1}^{{N_t}} {\left| {{\bf{h}}_k^H{{\bf{w}}_{r,i}}} \right|}^2  + {\sigma ^2}} }} \ge {r_{k,{\rm{th}}}},k \in {\cal K},\label{P1_const1}\\
		&\qquad {\rm{tr}}\left( {{\bf{AR}}{{\bf{A}}^H}{{\left( {{\bf{BR}}{{\bf{B}}^H} + {\sigma ^2}{{\bf{I}}_{{N_r}}}} \right)}^{ - 1}}} \right) \ge {r_{r,{\rm{th}}}},\label{P1_const2}\\
		& \qquad \sum\limits_{l = 1}^{L - 1} {\sum\limits_{j = l + 1}^L {{{\left| {{\bf{a}}_t^H\left( {{\theta _l}} \right)\left( {\sum\limits_{k = 1}^K {{{\bf{w}}_{c,k}}{\bf{w}}_{c,k}^H + } \sum\limits_{i = 1}^{{N_t}} {{{\bf{w}}_{r,i}}{\bf{w}}_{r,i}^H} } \right){\bf{a}}_t\left( {{\theta _j}} \right)} \right|}^2}} }  \le  {\epsilon_{{\rm{th}}}},  \label{P1_const3}\\
		&\qquad \left| {{v_m}} \right| = 1,m \in {\cal M},  \label{P1_const4}
	\end{align}
\end{subequations}
where    constraint  \eqref{P1_const1} denotes  the minimum  SINR, i.e., $r_{k,{\rm th}}$,   required by user $k, k\in {\cal K}$; constraint  \eqref{P1_const2} represents that the received radar SINR should exceed the minimum threshold $r_{r,{\rm th}}$;  constraint   \eqref{P1_const3} denotes that the cross-correlation
 pattern between the probing signals
 at a number of given target locations must be  smaller than $\epsilon_{\rm th}$; 
constraint \eqref{P1_const4} denotes the unit-modulus 
constraint on each  IRS reflection coefficient.

Problem \eqref{P1} is non-convex  since the optimization variables are highly coupled in   constraints  \eqref{P1_const1}-\eqref{P1_const3} and the unit-modulus constraint is imposed on each reflection coefficient $v_m$ in \eqref{P1_const4},   there are no standard methods for solving such
non-convex optimization problem optimally in general. In the following, we further  study two cases, namely, case I and case II, based on the presence or absence of the  cross-correlation
pattern constraint \eqref{P1_const3} and the interference introduced by the IRS in Section III and Section IV, and then propose two algorithm, namely, a penalty-based algorithm and an SDR-based algorithm, to solve them, respectively.

%

 \section{Proposed Solution to  Case I}
In this section, we study problem \eqref{P1} by assuming that the interference introduced by the  IRS, i.e., ${{{\bf{G}}_r}{\bf{\Theta }}{{\bf{G}}_t}{\bf{s}}}$ in \eqref{echo_signals},  is perfectly canceled at the Radcom BS and ignoring  the cross-correlation pattern design. 
Thus, the radar SINR in \eqref{radarsinr} is reduced to ${\rm{SIN}}{{\rm{R}}_r} = {\rm{tr}}\left( {{\bf{AR}}{{\bf{A}}^H}} \right)/{\sigma ^2} = \left( {\sum\limits_{k = 1}^K {{\bf{w}}_{c,k}^H} {{\bf{A}}^H}{\bf{A}}{{\bf{w}}_{c,k}}{\rm{ + }}\sum\limits_{i = 1}^{{N_t}} {{\bf{w}}_{r,i}^H} {{\bf{A}}^H}{\bf{A}}{{\bf{w}}_{r,i}}} \right)/{\sigma ^2}$. 
Accordingly,  problem   \eqref{P1} can be simplified to 
 \begin{subequations} \label{P1_without}
	\begin{align}
	&\mathop {\min }\limits_{{\{v_m\}},{\left\{ {{{\bf{w}}_{c,i}}} \right\},\left\{ {{{\bf{w}}_{r,i}}} \right\}}} \sum\limits_{k = 1}^K {{{\left\| {{{\bf{w}}_{c,k}}} \right\|}^2} + } \sum\limits_{i = 1}^{{N_t}} {{{\left\| {{{\bf{w}}_{r,i}}} \right\|}^2}} \\
	& {\rm s.t.}~ \sum\limits_{k = 1}^K {{\bf{w}}_{c,k}^H} {{\bf{A}}^H}{\bf{A}}{{\bf{w}}_{c,k}}{\rm{ + }}\sum\limits_{i = 1}^{{N_t}} {{\bf{w}}_{r,i}^H} {{\bf{A}}^H}{\bf{A}}{{\bf{w}}_{r,i}} \ge {r_{r,{\rm{th}}}}{\sigma ^2}, \\
	&\qquad \eqref{P1_const1},\eqref{P1_const4}.
	\end{align}
\end{subequations}
It is not difficult to check  that problem \eqref{P1_without} is  non-convex.  In the following, we first exploit the hidden structure of problem \eqref{P1_without} and derive  the following theorem:

\textbf{\emph{Theorem 1:}}
Under the assumptions  of independently distributed  complex amplitudes of targets, i.e., the non-zero singular values of ${\bf A}^H{\bf A}$ are not same, and    amplitudes  of targets and user channels are uncorrelated,  the optimal solution of problem \eqref{P1_without} satisfies $ {{{\mathbf{w}}_{r,i}^{\rm opt}}}={\bf 0}, i\in {{\cal N}_t}$.

\hspace*{\parindent}\textit{Proof}: Please refer to Appendix~A. \hfill\rule{2.7mm}{2.7mm}

Theorem 1 indicates that the  dedicated   radar  beams, i.e., $\left\{ {{{\mathbf{w}}_{r,i}}}, i\in {{\cal N}_t} \right\}$, are not needed for achieving the minimum   Radcom BS transmit power. This 
can be intuitively understood since sending dedicated radar signals not only consumes
transmit power but also potentially causes interference to communication users. 
Based on Theorem 1,  the  implementation complexity  of the Radcom BS as well as   the algorithm design is reduced.

To obtain a high-quality solution, a  penalty-based algorithm is proposed to decouple the constraint coupling between the variables in different blocks. 
Specifically, we first introduce several auxiliary variables $\left\{ {x_{k,i}^c,{\mathbf{y}}_i^c,k \in {\cal K},i \in {\cal K}} \right\}$, and define ${\bf{h}}_k^H{{\bf{w}}_{c,i}} = x_{k,i}^c$ and  ${\bf{A}}{{\bf{w}}_{c,i}} = {\bf{y}}_i^c$, $k \in {\cal K},i \in {\cal K}$, problem \eqref{P1_without} (by dropping radar beams) can be rewritten as 
 \begin{subequations} \label{P2}
	\begin{align}
	&\mathop {\min }\limits_{\{ {v_m}\} ,\left\{ {{{\mathbf{w}}_{c,i}}} \right\},\left\{ {x_{k,i}^c,{\mathbf{y}}_i^c} \right\}} \sum\limits_{k = 1}^K {{{\left\| {{{\mathbf{w}}_{c,k}}} \right\|}^2}}   \label{P2_obj} \\
	& {\rm s.t.}~\frac{{{{\left| {x_{k,k}^c} \right|}^2}}}{{\sum\limits_{i \ne k}^K {{{\left| {x_{k,i}^c} \right|}^2} + {\sigma ^2}} }} \ge {r_{k,{\text{th}}}},k \in {\cal K},\label{P2_const1}\\
	&\qquad  \sum\limits_{i = 1}^K {{{\left\| {{\bf{y}}_i^c} \right\|}^2}}  \ge {{ \sigma} ^2}{r_{r,{\rm{th}}}},\label{P2_const2}\\
		&\qquad {\bf{h}}_k^H{{\bf{w}}_{c,i}} = x_{k,i}^c, {\bf{A}}{{\bf{w}}_{c,i}} = {\bf{y}}_i^c, i\in {\cal K},k\in {\cal K},\label{P2_const3}\\
	&\qquad \eqref{P1_const4}.
	\end{align}
\end{subequations}
 It can be seen that  the optimization variables in constraints
\eqref{P2_const1} and \eqref{P2_const2} are fully decoupled since these two constraints do not contain any common optimization variables. We then use \eqref{P2_const3}    as penalty terms that are added to the objective function \eqref{P2_obj}, yielding the following penalty-based optimization problem
 \begin{subequations} \label{P2_penalty}
	\begin{align}
	&\mathop {\min }\limits_{\{ {v_m}\} ,\left\{ {{{\mathbf{w}}_{c,i}}} \right\},{\left\{ {x_{k,i}^c,{\mathbf{y}}_i^c} \right\}} } \sum\limits_{k = 1}^K {{{\left\| {{{\mathbf{w}}_{c,k}}} \right\|}^2} + } \frac{1}{{2\rho }}\left( {\sum\limits_{k = 1}^K {\sum\limits_{i = 1}^K {{{\left| {{\mathbf{h}}_k^H{{\mathbf{w}}_{c,i}} - x_{k,i}^c} \right|}^2} + \sum\limits_{i = 1}^K {{{\left\| {{\mathbf{A}}{{\mathbf{w}}_{c,i}} - {\mathbf{y}}_i^c} \right\|}^2}} } } } \right)\label{P2_penalty_obj}\\
	& {\rm s.t.}~\eqref{P1_const4},\eqref{P2_const1},\eqref{P2_const2}.
	\end{align}
\end{subequations}
where $\rho$ $(\rho>0)$ represents the penalty coefficient used to penalize the violation of the equality in  constraint  \eqref{P2_const3}.  By gradually decreasing the value of $\rho$  over outer layer iterations,  as $\rho  \to 0$, it follows that ${1}/({{2\rho }}) \to \infty $. As such, the equality in  \eqref{P2_const3}   is guaranteed by   the optimal solution to problem \eqref{P2_penalty}. With fixed $\rho$, it can be seen that problem \eqref{P2_penalty} is still non-convex. To tackle this non-convex optimization problem, we divide all the optimization variables into three  blocks in the inner layer, namely, 1) transmit beamformers $\left\{ {{{\bf{w}}_{c,i}}} \right\}$, 2) IRS phase shifts $\{v_m\}$, and 3) auxiliary variables $\left\{ {x_{k,i}^c,{\bf{y}}_i^c} \right\}$, and then
alternately optimize each block, until convergence is achieved.
\subsection{Inner Layer Optimization}
In this subsection, we elaborate on how to solve    the above three subproblems.  
In particular, we obtain a closed-form and/or a semi-closed-form solution to each of these three subproblems.
\subsubsection{For any given phase shifts $\{v_m\}$ and auxiliary variables $\left\{ {x_{k,i}^c,{\bf{y}}_i^c} \right\}$, the subproblem corresponding to the transmit beamformer optimization is given by}
\begin{align} \label{P2_penalty_1}
	\mathop {\min }\limits_{\left\{ {{{\mathbf{w}}_{c,i}}} \right\}} \sum\limits_{k = 1}^K {{{\left\| {{{\mathbf{w}}_{c,k}}} \right\|}^2} + } \frac{1}{{2\rho }}\left( {\sum\limits_{k = 1}^K {\sum\limits_{i = 1}^K {{{\left| {{\mathbf{h}}_k^H{{\mathbf{w}}_{c,i}} - x_{k,i}^c} \right|}^2} + \sum\limits_{i = 1}^K {{{\left\| {{\mathbf{A}}{{\mathbf{w}}_{c,i}} - {\mathbf{y}}_i^c} \right\|}^2}} } } } \right)
\end{align}
It can be readily observed that problem \eqref{P2_penalty_1} is a convex quadratic minimization problem without constraints. Thus, we can obtain its optimal solution by exploiting  the first-order optimality conditions. 
Specifically,  by taking the first-order derivative of the objective function \eqref{P2_penalty_1} with respect to (w.r.t.)  ${{{\bf{w}}_{c,i}}}$  and setting it to zero, the closed-form solution of ${{{\bf{w}}_{c,i}}}$ can be obtained as 
\begin{align}
{\bf{w}}_{c,i}^{{\rm{opt}}} = \frac{1}{{2\rho }}{\left( {{{\bf{I}}_{{N_t}}} + \frac{1}{{2\rho }}\left( {\sum\limits_{k = 1}^K {{{\bf{h}}_k}{\bf{h}}_k^H + {{\bf{A}}^H}{\bf{A}}} } \right)} \right)^{ - 1}} \left( {\sum\limits_{k = 1}^K {{{\bf{h}}_k}x_{k,i}^c + {{\bf{A}}^H}{\bf{y}}_i^c} } \right),  i\in {\cal K}. \label{optimalbeamform1}
\end{align}
\subsubsection{For any given transmit beamformers $\{{\bf w}_{c,i}\}$ and auxiliary variables $\left\{ {x_{k,i}^c,{\bf{y}}_i^c} \right\}$, the subproblem corresponding to the IRS phase-shift optimization is given by (by dropping constants $\rho$, ${{{\left\| {{{\bf{w}}_{c,i}}} \right\|}^2}}$'s, and ${{{\left\| {{\bf{A}}{{\bf{w}}_{c,i}} - {\bf{y}}_i^c} \right\|}^2}}$'s)}
 \begin{subequations} \label{P2_penalty_2}
	\begin{align}
	&\mathop {\min }\limits_{\{ {v_m}\} } \sum\limits_{k = 1}^K {\sum\limits_{i = 1}^K {{{\left| {{\mathbf{h}}_k^H{{\mathbf{w}}_{c,i}} - x_{k,i}^c} \right|}^2}} }  \label{P2_penalty_2_obj} \\
	&{\rm s.t.}~\eqref{P1_const4}.
	\end{align}
\end{subequations}
 Recall that  ${\bf{h}}_k^H{\rm{ = }}{\bf{h}}_{d,k}^H + {\bf{h}}_{r,k}^H{\bf{\Theta }}{{\bf{G}}_t},k \in {\cal K}$,  it is not difficult to verify  that objective function  \eqref{P2_penalty_2_obj} is a    convex quadratic function. However, due to the unit-modulus constraint on each IRS phase shift in \eqref{P1_const4}, problem \eqref{P2_penalty_2} is a non-convex optimization problem.
To solve this problem, an element-wise algorithm is proposed, where the main idea behind it is to optimize one
phase shift  with the other phase shifts are fixed. 
Specifically,  we  rewrite  ${\bf{h}}_k^H{\rm{ = }}{\bf{h}}_{d,k}^H + {\bf{h}}_{r,k}^H{\bf{\Theta }}{{\bf{G}}_t}$   as 
\begin{align}
{\bf{h}}_k^H{\rm{ = }}{\bf{h}}_{d,k}^H{\rm{ + }}{{\bf{v}}^H}{\rm{diag}}\left( {{\bf{h}}_{r,k}^H} \right){{\bf{G}}_t}, k \in {\cal K},
\end{align}
where ${{\bf{v}}^H} = \left[ {{v_1}, \ldots ,{v_M}} \right]$. Then, define ${\bf{q}}_{k,i}^c{\rm{ = diag}}\left( {{\bf{h}}_{r,k}^H} \right){{\bf{G}}_t}{{\bf{w}}_{c,i}}$, $k\in{\cal K}, i\in{\cal K}$,  we can write ${{{\left| {{\bf{h}}_k^H{{\bf{w}}_{c,i}} - x_{k,i}^c} \right|}^2}}$ w.r.t. the $m$th IRS phase shift, i.e., $v_m$, in a more compact form given by 
\begin{align}
{\left| {{\bf{h}}_k^H{{\bf{w}}_{c,i}} - x_{k,i}^c} \right|^2} \overset{(a)}{= }{\left| {{{\left[ {{\bf{q}}_{k,i}^c} \right]}_m}} \right|^2} + 2{\mathop{\rm Re}\nolimits} \left\{ {{v_m}{{\left[ {{\bf{q}}_{k,i}^c} \right]}_m}{a_{k,i,\bar m}^{c,H}}} \right\}+{\left| {a_{k,i,\bar m}^{c}} \right|^2}, k \in{\cal K},i\in{\cal K},
\end{align}
where  $(a)$ holds due to $\left| {{v_m}} \right| = 1,\forall m$, and $a_{k,i,\bar m}^c = \sum\limits_{j \ne m}^M {{v_j}{{\left[ {{\bf{q}}_{k,i}^c} \right]}_j} + {\bf{h}}_{d,k}^H{{\bf{w}}_{c,i}} - x_{k,i}^c}$.

Therefore,  problem \eqref{P2_penalty_2} regarding  to  the $m$th IRS phase-shift optimization  becomes (by dropping irrelevant constants w.r.t. $v_m$)
 \begin{subequations} \label{P2_penalty_2_mth}
	\begin{align}
	&\mathop {\min }\limits_{{v_m}} {\text{Re}}\left\{ {{v_m}\left( {\sum\limits_{k = 1}^K {\sum\limits_{i = 1}^K {{{\left[ {{\mathbf{q}}_{k,i}^c} \right]}_m}a_{k,i,\bar m}^{c,H}} } } \right)} \right\}\\
	&{\rm s.t.}~\left| {{v_m}} \right| = 1.  \label{P2_penalty_2_const5_mth}
	\end{align}
\end{subequations}
It can be  observed that the objective function of problem \eqref{P2_penalty_2_mth} is  linear w.r.t. $v_m$, and the  optimal solution to problem \eqref{P2_penalty_2_mth} can be obtained as 
\begin{align}
v_m^{{\text{opt}}} =  - \exp \left( {j\arg {{\left( {\sum\limits_{k = 1}^K {\sum\limits_{i = 1}^K {{{\left[ {{\mathbf{q}}_{k,i}^c} \right]}_m}a_{k,i,\bar m}^c} } } \right)}^*}} \right). \label{P2_optimal_phase}
\end{align}
Based on \eqref{P2_optimal_phase}, we can alternately optimize each IRS phase shift in an iterative manner. 
\subsubsection{For any given IRS phase shifts $\{v_m\}$ and  transmit beamformers $\{{\bf w}_{c,i}\}$, the auxiliary variables can be optimized by solving the following subproblem (by dropping constants $\rho$ and ${{{\left\| {{{\bf{w}}_{c,i}}} \right\|}^2}}$'s)}

\begin{subequations} \label{P2_penalty_3}
	\begin{align}
	&\mathop {\min }\limits_{\{ x_{k,i}^c\} ,\{ {\mathbf{y}}_i^c\} } \sum\limits_{k = 1}^K {\sum\limits_{i = 1}^K {{{\left| {{\mathbf{h}}_k^H{{\mathbf{w}}_{c,i}} - x_{k,i}^c} \right|}^2} + } } \sum\limits_{i = 1}^K {{{\left\| {{\mathbf{A}}{{\mathbf{w}}_{c,i}} - {\mathbf{y}}_i^c} \right\|}^2}}   \\
	& {\rm s.t.}~\eqref{P2_const1},\eqref{P2_const2}.
	\end{align}
\end{subequations}
Since optimization variables w.r.t. different blocks $\left\{ {x_{k,i}^c}, i\in {\cal K} \right\}$ for $ k \in {\cal K}$ and $\left\{ {{\bf{y}}_i^c},i \in {\cal K} \right\}$ are separated in both the objective function and constraints. Therefore, problem \eqref{P2_penalty_3} can be divided into  $K+1$ separated subproblems, which can be solved in a parallel manner as follows. 

On the one hand, the subproblem regarding to   the $k$th block $\left\{ {x_{k,i}^c}, i\in {\cal K} \right\}$ is given by 
\begin{subequations} \label{P2_penalty_31}
	\begin{align}
	&\mathop {\min }\limits_{\{ x_{k,i}^c\} } \sum\limits_{i = 1}^K {{{\left| {{\mathbf{h}}_k^H{{\mathbf{w}}_{c,i}} - x_{k,i}^c} \right|}^2}}   \\
	& {\rm s.t.}~\eqref{P2_const1}.
	\end{align}
\end{subequations}
It is not difficult to see that problem \eqref{P2_penalty_31} is a quadratically constrained quadratic program (QCQP) problem with convex objective function and non-convex constraint \eqref{P2_const1}. Fortunately, it was shown in \cite[Appendix B.1]{boyd2004convex} that the strong duality holds for any optimization problem with quadratic objective and one quadratic inequality constraint, provided Slater’s condition holds. This result shows that the optimal solution to problem \eqref{P2_penalty_31} can be obtained by solving its dual problem.
By introducing dual variable ${\lambda _{2,k}}$ $\left( {{\lambda _{2,k}} \ge 0} \right)$  associated with constraint \eqref{P2_const1}, the Lagrangian function of problem \eqref{P2_penalty_31} is given by 
\begin{align}
{{\cal L}_2}\left( {x_{k,i}^c,{\lambda _{2,k}}} \right){\text{ }} = \sum\limits_{i = 1}^K {{{\left| {{\mathbf{h}}_k^H{{\mathbf{w}}_{c,i}} - x_{k,i}^c} \right|}^2}}  + {\lambda _{2,k}}\left( {{r_{k,{\text{th}}}}\left( {\sum\limits_{i \ne k}^K {{{\left| {x_{k,i}^c} \right|}^2} + {\sigma ^2}} } \right) - {{\left| {x_{k,k}^c} \right|}^2}} \right). \label{Lagrangian}
\end{align}
Accordingly, the corresponding dual function is given by $f_2\left( {{\lambda _{2,k}}} \right) = \mathop {\min }\limits_{x_{k,i}^c} {\cal L}_2\left( {x_{k,i}^c,{\lambda _{2,k}}} \right)$.

\textbf{\emph{Lemma 2:}}
To make dual function  $f_2\left( {{\lambda _{2,k}}} \right)$ bounded, we must have 
\begin{align}
0 \le {\lambda _{2,k}} < 1.
\end{align} 

\hspace*{\parindent}\textit{Proof}: Please refer to Appendix~B. \hfill\rule{2.7mm}{2.7mm}

Based on Lemma~$2$, the optimal solution to ${f_2}\left( {{\lambda _{2,k}}} \right)$ can be obtained by leveraging the first-order optimality conditions. Specifically,  
by taking the first-order derivative of ${\cal L}_2\left( {x_{k,i}^c,{\lambda _{2,k}}} \right)$ w.r.t. ${x_{k,i}^c}$ and setting it to zero, we obtain the optimal solution as
\begin{align}
x_{k,i}^{c,{\rm{opt}}}\left( {{\lambda _{2,k}}} \right) = \left\{ \begin{array}{l}
\frac{{{\bf{h}}_k^H{{\bf{w}}_{c,i}}}}{{1 + {\lambda _{2,k}}{r_{k,{\rm{th}}}}}},i \ne k, i \in {\cal K},\\
\frac{{{\bf{h}}_k^H{{\bf{w}}_{c,k}}}}{{1-{\lambda _{2,k}} }},i = k.
\end{array} \right.
\end{align}	
Recall that for the optimal solutions $x_{k,i}^{c,{\rm{opt}}}\left( {{\lambda _{2,k}}} \right)$ and ${{\lambda _{2,k}^{\rm opt}}}$, the following complementary slackness condition must be satisfied \cite{boyd2004convex}	
\begin{align}
\lambda _{2,k}^{{\text{opt}}}\left( {{r_{k,{\text{th}}}}\left( {\sum\limits_{i \ne k}^K {{{\left| {x_{k,i}^{c,{\text{opt}}}\left( {\lambda _{2,k}^{{\text{opt}}}} \right)} \right|}^2} + {\sigma ^2}} } \right) - {{\left| {x_{k,k}^{c,{\text{opt}}}\left( {\lambda _{2,k}^{{\text{opt}}}} \right)} \right|}^2}} \right) = 0.
\end{align}
Next, we    check whether $\lambda _{2,k}^{{\rm{opt}}}=0$ is the optimal solution or not. If
\begin{align}
{r_{k,{\text{th}}}}\left( {\sum\limits_{i \ne k}^K {{{\left| {x_{k,i}^{c,{\text{opt}}}\left( 0 \right)} \right|}^2} + {\sigma ^2}} } \right) - {\left| {x_{k,k}^{c,{\text{opt}}}\left( 0 \right)} \right|^2} < 0, 
\end{align}	
which indicates that the optimal dual variable	$\lambda _{2,k}^{{\rm{opt}}}$ equals to $0$, otherwise, the optimal
$\lambda _{2,k}^{{\rm{opt}}}$ is a positive value, i.e., ${\lambda _{2,k}^{{\rm{opt}}}}>0$, and  should satisfy 
\begin{align}
 {r_{k,{\text{th}}}}\left( {\sum\limits_{i \ne k}^K {{{\left| {x_{k,i}^{c,{\text{opt}}}\left( {\lambda _{2,k}^{{\text{opt}}}} \right)} \right|}^2} + {\sigma ^2}} } \right) - {\left| {x_{k,k}^{c,{\text{opt}}}\left( {\lambda _{2,k}^{{\text{opt}}}} \right)} \right|^2} = 0.
\end{align}
It can be seen    that   ${\left| {x_{k,i}^{c,{\rm{opt}}}\left( {\lambda _{2,k}} \right)} \right|}$ for $i\ne k$ is   monotonically decreasing  with ${\lambda _{2,k}}$, while $\left| {x_{k,k}^{c,{\rm{opt}}}\left( {\lambda _{2,k}} \right)} \right|$ is  monotonically increasing  with ${\lambda _{2,k}}$ for $0 < \lambda _{2,k} < 1$. As such, the optimal $\lambda _{2,k}^{{\rm{opt}}}$ can be obtained by applying a simple bisection search method.

On the other hand, the subproblem regarding to     block $\left\{ {{\bf{y}}_i^c},i \in {\cal K},  \right\}$ is formulated as 
\begin{subequations} \label{P2_penalty_32}
	\begin{align}
	&\mathop {\min }\limits_{\{ {\mathbf{y}}_i^c\} } \sum\limits_{i = 1}^K {{{\left\| {{\mathbf{A}}{{\mathbf{w}}_{c,i}} - {\mathbf{y}}_i^c} \right\|}^2}}  \\
	&{\rm s.t.}~ \eqref{P2_const2}.
	\end{align}
\end{subequations}
It can be observed that problem \eqref{P2_penalty_32} is also a  QCQP problem with a quadratic objective and one quadratic inequality constraint.  Following  \cite[Appendix B.1]{boyd2004convex},  the strong duality also holds for problem \eqref{P2_penalty_32}. 
Thus,   the optimal solution to problem \eqref{P2_penalty_32} can be obtained by solving its dual problem.
By introducing dual variable ${\lambda _3}$ $\left( {{\lambda _3} \ge 0} \right)$  associated with constraint \eqref{P2_const2}, the Lagrangian function of \eqref{P2_penalty_32} is given by
\begin{align}
{{\cal L}_3}\left( {{\mathbf{y}}_i^c,{\lambda _3}} \right){\text{ = }}\sum\limits_{i = 1}^K {{{\left\| {{\mathbf{A}}{{\mathbf{w}}_{c,i}} - {\mathbf{y}}_i^c} \right\|}^2}} {\text{ + }}{\lambda _3}\left( {{{ \sigma }^2}{r_{r,{\text{th}}}} - \sum\limits_{i = 1}^K {{{\left\| {{\mathbf{y}}_i^c} \right\|}^2}} } \right). \label{P2_penalty_32_larg}
\end{align}
Let  $ f_3\left( {\lambda _3}  \right) = \mathop {\min }\limits_{{\bf{y}}_i^c} { {\cal L}_3}\left( {{\bf{y}}_i^c,{\lambda _3} } \right)$ be  the dual function of problem \eqref{P2_penalty_32}, we have the following lemma:

\textbf{\emph{Lemma 3:}}
	To guarantee the dual function $f_3\left( {\lambda _3} \right) $ be bounded, it follows that  
\begin{align}
0 \le {\lambda _3} < 1.
\end{align}

\hspace*{\parindent}\textit{Proof}: The proof is similar to Lemma~$2$ and is omitted here for brevity. \hfill\rule{2.7mm}{2.7mm}

Based on Lemma~$3$,
by exploiting the first-order optimality conditions, the optimal solution to ${f_3}\left( {{\lambda _3}} \right)$ is given by
\begin{align}
{\bf{y}}_i^{c,{\rm{opt}}}({\lambda _3} ) = \frac{{{\bf{A}}{{\bf{w}}_{c,i}}}}{{1 - {\lambda _3}}},i \in {\cal K}.\label{P2_penalty_opt3}
\end{align}
The optimal dual variable ${\lambda _3}^{\rm opt}$ should be chosen for ensuring that the following complementary
slackness condition is satisfied:
\begin{align}
\lambda _3^{\rm opt}\left( {{{ \sigma }^2}{r_{r,{\text{th}}}} - \sum\limits_{i = 1}^K {{{\left\| {{\mathbf{y}}_i^c\left( {\lambda _3^{{\rm{opt}}}} \right)} \right\|}^2}} } \right){\text{ = }}0. \label{P2_penalty_complementy2}
\end{align}
Define $\Gamma \left( {\lambda _3} \right) = \sum\limits_{i = 1}^K {{{\left\| {{\bf{y}}_i^c\left( {\lambda _3} \right)} \right\|}^2}} $. Then, substituting  \eqref{P2_penalty_opt3} into $\Gamma \left( {\lambda _3}  \right)$, we have
\begin{align}
\Gamma \left( {{\lambda _3}} \right) = \frac{{\sum\limits_{i = 1}^K {{{\left\| {{\mathbf{A}}{{\mathbf{w}}_{c,i}}} \right\|}^2}} }}{{{{\left( {1 - {\lambda _3}} \right)}^2}}}.\label{P2_penalty_32_Gmama}
\end{align}
It can be observed that $\Gamma \left( {\lambda _3}  \right)$ is monotonically increasing with ${\lambda _3}$ for $0\le{\lambda _3}<1$. Thus,
 if ${{ \sigma} ^2}{r_{r,{\rm{th}}}}{\rm{ - }}\Gamma \left( 0 \right) < 0$, which indicates that   the optimal
dual variable is ${{\lambda _3} ^{{\rm{opt}}}}=0$.  Otherwise,
the optimal ${{\lambda _3} ^{{\rm{opt}}}}$ can be obtained by solving  the following equation:
\begin{align}
\Gamma \left( {\lambda _3}  \right)={{ \sigma} ^2}{r_{r,{\rm{th}}}}.\label{P2_penalty_32_equ}
\end{align}
By exploiting the   monotonic   property of $\Gamma \left( {\lambda _3} \right)$, the solution ${{\lambda _3} ^{{\rm{opt}}}}$ that  satisfies \eqref{P2_penalty_32_equ}  can be readily obtained by applying the simple bisection method  searching  from  $0$ to $1$ .
\subsection{Outer Layer Update}
In the outer layer, we need to gradually decrease  the penalty coefficient $\rho^{t}$ in  the $t$th iteration, which can be updated as follow 
\begin{align}
{\rho ^{t}}{\rm{ = }}c{\rho ^{t-1}},\label{updatecoefficient}
\end{align}
where $c$ $(0<c<1)$ denotes the updated step size.  Generally, a larger value of $c$ can achieve better performance but at the
cost of more iterations for updating in the outer layer. Although  a smaller value of $c$ requires less  outer layer iterations for updating, the penalty algorithm  is more easily diverged. From empirical test, it is promising to choose $c$ from $0.7$ to $0.9$ to balance the system performance and computational complexity. 
\subsection{Overall Algorithm}
Next, we provide  the termination condition for our proposed algorithm, which is given by 
\begin{align}
\xi  = \max \left\{ {{{\left| {{\bf{h}}_k^H{{\bf{w}}_{c,i}} - x_{k,i}^c} \right|}^2},{{\left\| {{\bf{A}}{{\bf{w}}_{c,i}} - {\bf{y}}_i^c} \right\|}_\infty^2 },i \in {\cal K},k \in {\cal K}} \right\}, \label{terminationindicator}
\end{align}
where $\xi$ denotes the termination indicator. If $\xi$ is smaller than a predefined value, which indicates that  constraint \eqref{P2_const3}  is   met with equality.  The details of the proposed penalty-based  algorithm are summarized in Algorithm~\ref{alg1}.
\begin{algorithm}[!t]
	\caption{Penalty-based algorithm for solving problem \eqref{P1_without}.}
	\label{alg1}
	\begin{algorithmic}[1]
		\STATE  \textbf{Initialize} ${\bf v}$, $x_{k,i}^{c},{\bf{y}}_i^{c}$, $c$, $\rho^t$, $\varepsilon_1$, and   $\varepsilon_2$.
		\STATE  \textbf{repeat: outer layer}
		\STATE \quad \textbf{repeat: inner layer }
		\STATE  \qquad Update transmit beamformers $\{{\bf w}_{c,i}\}$ based on \eqref{optimalbeamform1}.
		\STATE  \qquad Update IRS phase shifts $\{v_m\}$  based on \eqref{P2_optimal_phase}.
		\STATE  \qquad Update auxiliary variables $\{x_{k,i}^{c}\}$ by solving problem \eqref{P2_penalty_31}.
		\STATE  \qquad Update auxiliary variables $\{{\bf{y}}_i^{c}\}$ by solving problem \eqref{P2_penalty_32}.
		\STATE \quad \textbf{until} the fractional decrease of the objective value of \eqref{P2_penalty} is below a threshold $\varepsilon_1$.
		\STATE  \quad Update penalty coefficient $\rho$ based on \eqref{updatecoefficient}.
		\STATE \textbf{until} termination indicator  $\xi$ is below a predefined threshold $\varepsilon_2$.
	\end{algorithmic}
\end{algorithm}
In Algorithm~\ref{alg1},  each block in the inner layer  is optimally  solved and there is no coupling between  the variables in different blocks. Following 
\cite[Theorem 4.1]{shi2016joint}, the solution obtained by   Algorithm~\ref{alg1} is guaranteed to converge to a stationary point.


The computational complexity of Algorithm~\ref{alg1} is calculated as follows.   In steps $4$ and $5$,  the  closed-form  solutions are obtained, whose computational complexity are given by ${\cal O}\left( {N_t^3K} \right)$ and ${\cal O}\left( {MN_r^2} \right)$, respectively. In steps $6$ and $7$, a bisection method is applied, whose computational complexity are given by  ${\cal O}\left( {K{{\log }_2}\left( {\frac{1}{{{\varepsilon _3}}}} \right)N_r^2} \right)$ and ${\cal O}\left( {{{\log }_2}\left( {\frac{1}{{{\varepsilon _3}}}} \right)N_r^2} \right)$, respectively, where ${{\varepsilon _3}}$ denotes the iteration accuracy. Therefore, the  overall complexity of  Algorithm~\ref{alg1} is given by ${\cal O}\left( {{I_{{\rm{outer}}}}\left( {{I_{{\rm{inner}}}}\left( {N_t^3K + MN_r^2 + \left( {K + 1} \right){{\log }_2}\left( {\frac{1}{{{\varepsilon _3}}}} \right)N_r^2} \right)} \right)} \right)$, where ${{I_{{\rm{inner}}}}}$ and ${{I_{{\rm{outer}}}}}$ denote the number of iterations required for  convergence in the inner layer and outer layer, respectively.

 \section{Proposed Solution to  Case II}
In this section, we consider  case II  where the interference introduced by  IRS  is  uncanceled and the  cross-correlation pattern design is required.  The  penalty-based algorithm proposed in  case I is not applicable to this case. To tackle this difficulty, an SDR-based AO algorithm  is proposed.

Recall that ${{\mathbf{Z}}_r} = {{\mathbf{W}}_r}{\mathbf{W}}_r^H$ and   ${{\bf{W}}_{c,k}} = {{\bf{w}}_{c,k}}{\bf{w}}_{c,k}^H$ defined in Appendix A, which satisfy ${\bf Z}_r\succeq{\bf 0}$, ${{\bf{W}}_{c,k}} \succeq {\bf{0}}$, and ${\rm{rank}}\left( {{{\bf{W}}_{c,k}}} \right) = 1,k \in {\cal K}$.  We can rewrite ${\bf{R}}$ defined in \eqref{radarsinr} as ${\bf{R}} = \sum\limits_{k = 1}^K {{{\bf{W}}_{c,k}}}  + {{\bf{Z}}_r}$.
Since the rank-one constraint is non-convex, we apply SDR to relax
this constraint. As a result, the SDR of problem \eqref{P1} is given by  
 \begin{subequations} \label{P3}
	\begin{align}
		&\mathop {\min }\limits_{\{ {v_m}\} ,\left\{ {{{\mathbf{W}}_{c,k}}} \right\},{{\mathbf{Z}}_r}} \sum\limits_{k = 1}^K {{\text{tr}}} \left( {{{\mathbf{W}}_{c,k}}} \right) + {\text{tr}}\left( {{{\mathbf{Z}}_r}} \right) \label{P3_obj} \\
		& {\rm s.t.}~{\mathbf{h}}_k^H\left( {\sum\limits_{k = 1}^K {{{\mathbf{W}}_{c,k}}}  + {{\mathbf{Z}}_r}} \right){{\mathbf{h}}_k} + {\sigma ^2} \le  \left( {\frac{1}{{{r_{k,{\text{th}}}}}} + 1} \right){\mathbf{h}}_k^H{{\mathbf{W}}_{c,k}}{{\mathbf{h}}_k},k \in {\cal K},\label{P3_const1}\\
& {\rm{tr}}\left( {{\bf{A}}\left( {\sum\limits_{k = 1}^K {{{\bf{W}}_{c,k}}}  + {{\bf{Z}}_r}} \right){{\bf{A}}^H}{{\left( {{\bf{B}}\left( {\sum\limits_{k = 1}^K {{{\bf{W}}_{c,k}}}  + {{\bf{Z}}_r}} \right){{\bf{B}}^H} + {\sigma ^2}{{\bf{I}}_{{N_r}}}} \right)}^{ - 1}}} \right) \ge {r_{r,{\rm{th}}}},\label{P3_const2_temp}\\
		& \qquad \sum\limits_{l = 1}^{L - 1} {\sum\limits_{j = l + 1}^L {{{\left| {{\bf{a}}_t^H\left( {{\theta _l}} \right)\left( {\sum\limits_{k = 1}^K {{{\bf{W}}_{c,k}}}  + {{\bf{Z}}_r}} \right){{\bf{a}}_t}\left( {{\theta _j}} \right)} \right|}^2}} }   \le  {\epsilon_{{\rm{th}}}},  \label{P3_const3}\\
		&\qquad \eqref{P1_const4}.
	\end{align}
\end{subequations}
It is not difficult to verify  that    constraints  \eqref{P1_const4},  \eqref{P3_const1},  and \eqref{P3_const2_temp} are all non-convex, which in general there are no efficient approaches to solve it optimally. In the following, we first derive the lower bound of   constraint \eqref{P3_const2_temp} based on  the identity  
\begin{align}
{\rm{tr}}\left( {{\bf{AR}}{{\bf{A}}^H}{{\left( {{\bf{BR}}{{\bf{B}}^H} + {\sigma ^2}{{\bf{I}}_{{N_r}}}} \right)}^{ - 1}}} \right) \ge \frac{{{\rm{tr}}\left( {{\bf{AR}}{{\bf{A}}^H}} \right)}}{{{\rm{tr}}\left( {{\bf{BR}}{{\bf{B}}^H} + {\sigma ^2}{{\bf{I}}_{{N_r}}}} \right)}}.
\end{align}
As such, constraint \eqref{P3_const2_temp} can be approximated as  a more tractable form  given by 
\begin{align}
 {\rm{tr}}\left( {{\bf{A}}\left( {\sum\limits_{k = 1}^K {{{\bf{W}}_{c,k}}}  + {{\bf{Z}}_r}} \right){{\bf{A}}^H}} \right) \ge {r_{r,{\rm{th}}}}\left( {{\rm{tr}}\left( {{\bf{B}}\left( {\sum\limits_{k = 1}^K {{{\bf{W}}_{c,k}}}  + {{\bf{Z}}_r}} \right){{\bf{B}}^H}} \right){\rm{ + }}{{ \sigma }^2}{N_r}} \right).\label{P3_const2}
\end{align}
 To tackle the non-convexity of   constraint \eqref{P1_const4}, we relax it as a convex form given by 
\begin{align}
 \left| {{v_m}} \right| \le 1,m \in {\cal M}. \label{P3_phase}
\end{align}
Then,  we   partition all optimization variables into two blocks,  i.e., transmit covariance matrices  $\left\{ {{{\mathbf{W}}_{c,k}},{{\mathbf{Z}}_r}} \right\}$ and  IRS phase shifts $\left\{ {v_m} \right\}$, and optimize these two blocks  in an iterative manner.
\subsection{Optimization of Transmit Covariance Matrices}
 For any given IRS phase shifts $\{v_m\}$, the subproblem regarding to the   transmit covariance matrix  optimization is given by 
 \begin{subequations} \label{P3_1}
	\begin{align}
		&\mathop {\min }\limits_{\left\{ {{{\mathbf{W}}_{c,k}}} \right\},{{\mathbf{Z}}_r}} \sum\limits_{k = 1}^K {{\text{tr}}} \left( {{{\mathbf{W}}_{c,k}}} \right) + {\text{tr}}\left( {{{\mathbf{Z}}_r}} \right)\label{P3_1_obj} \\
		& {\rm s.t.}~\eqref{P3_const1},\eqref{P3_const3},\eqref{P3_const2}.
	\end{align}
\end{subequations} 
It is not difficult to observe that the objective function as well as constraints are all convex, problem \eqref{P3_1} is thus convex and   can be solved by  the interior-point method \cite{boyd2004convex}.
 \subsection{Optimization of IRS phase shifts}
 For any given transmit covariance matrices  $\left\{ {{{\mathbf{W}}_{c,k}},{{\mathbf{Z}}_r}} \right\}$, the corresponding IRS phase-shift subproblem is given by 
  \begin{subequations} \label{P3_2}
 	\begin{align}
 	&{\rm{Find}}{\kern 1pt} {\kern 1pt} {\kern 1pt} {\kern 1pt} {\{v_m\}} \\
 	& {\rm s.t.}~\eqref{P3_const1},\eqref{P3_const2}, \eqref{P3_phase}.
 	\end{align}
 \end{subequations}
 Recall that ${\bf{h}}_k^H{\rm{ = }}{\bf{h}}_{d,k}^H{\rm{ + }}{{\bf{v}}^H}{\rm{diag}}\left( {{\bf{h}}_{r,k}^H} \right){{\bf{G}}_t}$, we can expand ${\bf{h}}_k^H{\bf{R}}{{\bf{h}}_k}$ as 
 \begin{align}
{\bf{h}}_k^H{\bf{R}}{{\bf{h}}_k}& = {\bf{h}}_{d,k}^H{\bf{R}}{{\bf{h}}_{d,k}} + 2{\mathop{\rm Re}\nolimits} \left\{ {{{\bf{v}}^H}{\rm{diag}}\left( {{\bf{h}}_{r,k}^H} \right){{\bf{G}}_t}{\bf{R}}{{\bf{h}}_{d,k}}} \right\} \notag\\
 &+ {{\bf{v}}^H}{\rm{diag}}\left( {{\bf{h}}_{r,k}^H} \right){{\bf{G}}_t}{\bf{RG}}_t^H{\rm{diag}}\left( {{{\bf{h}}_{r,k}}} \right){\bf{v}}\overset{\triangle}{=}{f_{1,k}}\left( {\bf{v}} \right), k \in {\cal K}. \label{P3_2_F1}
 \end{align}
  It is not difficult to see that ${f_{1,k}}\left( {\mathbf{v}} \right)$ is a quadratic function of $\bf v$, which is convex.
 
 Similarly,  we can rewrite ${\bf{h}}_k^H{{\bf{W}}_{c,k}}{{\bf{h}}_k}$ as 
 \begin{align}
{\bf{h}}_k^H{{\bf{W}}_{c,k}}{{\bf{h}}_k} &= {\bf{h}}_{d,k}^H{{\bf{W}}_{c,k}}{{\bf{h}}_{d,k}} + 2{\rm{Re}}\left\{ {{{\bf{v}}^H}{\rm{diag}}\left( {{\bf{h}}_{r,k}^H} \right){{\bf{G}}_t}{{\bf{W}}_{c,k}}{{\bf{h}}_{d,k}}} \right\} \notag\\
&+ {{\bf{v}}^H}{\rm{diag}}\left( {{\bf{h}}_{r,k}^H} \right){{\bf{G}}_t}{{\bf{W}}_{c,k}}{\bf{G}}_t^H{\rm{diag}}\left( {{{\bf{h}}_{r,k}}} \right){\bf{v}}\overset{\triangle}{=}{{f}_{2,k}}\left( {\mathbf{v}} \right), k \in {\cal K}.
 \end{align}
 Although ${{f}_{2,k}}\left( {\mathbf{v}} \right)$ is a quadratic function of $\bf v$,  the resulting set in \eqref{P3_const1} is
 not a convex set since the super-level set of a convex quadratic function is not convex in general. However, 
 we can linearize ${{\bf{v}}^H}{\rm{diag}}\left( {{\bf{h}}_{r,k}^H} \right){{\bf{G}}_t}{{\bf{W}}_{c,k}}{\bf{G}}_t^H{\rm{diag}}\left( {{{\bf{h}}_{r,k}}} \right){\bf{v}}$ into a
 linear form by taking its first-order Taylor expansion at any given point ${\bf v}^t$ in the $t$th iteration,  yielding the following inequality
\begin{align}
& {{\bf{v}}^H}{\rm{diag}}\left( {{\bf{h}}_{r,k}^H} \right){{\bf{G}}_t}{{\bf{W}}_{c,k}}{\bf{G}}_t^H{\rm{diag}}\left( {{{\bf{h}}_{r,k}}} \right){\bf{v}}\ge {\rm{ - }}{{\bf{v}}^{t,H}}{\rm{diag}}\left( {{\bf{h}}_{r,k}^H} \right){{\bf{G}}_t}{{\bf{W}}_{c,k}}{\bf{G}}_t^H{\rm{diag}}\left( {{{\bf{h}}_{r,k}}} \right){{\bf{v}}^t}\notag\\
&\qquad\qquad\qquad\qquad\qquad\quad+ 2{\mathop{\rm Re}\nolimits} \left\{ {{{\bf{v}}^{t,H}}{\rm{diag}}\left( {{\bf{h}}_{r,k}^H} \right){{\bf{G}}_t}{{\bf{W}}_{c,k}}{\bf{G}}_t^H{\rm{diag}}\left( {{{\bf{h}}_{r,k}}} \right){\bf{v}}} \right\}\overset{\triangle}{=}{\bar f}_{2,k}^{\rm lb}(\bf v).
\end{align}
As such, the lower bound of ${\bf{h}}_k^H{{\bf{W}}_{c,k}}{{\bf{h}}_k}$, denoted by ${f}_{2,k}^{{\rm{lb}}}\left( {\bf{v}} \right)$, is given by 
 \begin{align}
&{f}_{2,k}^{{\rm{lb}}}\left( {\bf{v}} \right)\overset{\triangle} {=} {\bf{h}}_{d,k}^H{{\bf{W}}_{c,k}}{{\bf{h}}_{d,k}} + 2{\rm{Re}}\left\{ {{{\bf{v}}^H}{\rm{diag}}\left( {{\bf{h}}_{r,k}^H} \right){{\bf{G}}_t}{{\bf{W}}_{c,k}}{{\bf{h}}_{d,k}}} \right\}+{\bar f}_{2,k}^{\rm lb}(\bf v),
 \label{P3_2_F2}
 \end{align} 
 which is a linear function of $\bf  v$ and  thus it is  convex. 
 
In addition, recall that ${\mathbf{R}} = \sum\limits_{k = 1}^K {{{\mathbf{W}}_{c,k}}}  + {{\mathbf{Z}}_r}$ and ${\bf{B}} = {{\bf{G}}_r}{\bf{\Theta G}}_t$, we can expand the right-hand side of \eqref{P3_const2}, i.e., ${{\rm{tr}}\left( {{\bf{B}}\left( {\sum\limits_{k = 1}^K {{{\bf{W}}_{c,k}}}  + {{\bf{Z}}_r}} \right){{\bf{B}}^H}} \right)}$, as 
\begin{align}
{\rm{tr}}\left( {{\bf{BR}}{{\bf{B}}^H}} \right)&{\rm{ = tr}}\left( {{{\bf{\Theta }}^H}{\bf{G}}_r^H{{\bf{G}}_r}{\bf{\Theta }}{{\bf{G}}_t}{\bf{RG}}_t^H} \right)
  = {{\bf{v}}^T}\left( {\left( {{\bf{G}}_r^H{{\bf{G}}_r}} \right) \odot {{\left( {{{\bf{G}}_t}{\bf{RG}}_t^H} \right)}^T}} \right){{\bf{v}}^*}. \label{functionofv}
\end{align}
It is not difficult to check that both  ${{\bf{G}}_r^H{{\bf{G}}_r}}$ and ${{{\bf{G}}_t}{\bf{RG}}_t^H}$  are positive semidefinite matrices,  the Hadamard product of  $ {{\bf{G}}_r^H{{\bf{G}}_r}} $ and ${\left( {{{\bf{G}}_t}{\bf{RG}}_t^H} \right)^T}$ is thus  a positive semidefinite matrix \cite{zhang2017matrix}. This indicates that ${\rm{tr}}\left( {{\bf{BR}}{{\bf{B}}^H}} \right)$ in \eqref{functionofv} is a quadratic function of ${\bf{v}}$, which is convex.

As a result, based on  \eqref{P3_2_F1}, \eqref{P3_2_F2}, and \eqref{functionofv},  the newly formulated problem is  given by 
  \begin{subequations} \label{P3_3}
	\begin{align}
	&{\rm{Find}}{\kern 1pt} {\kern 1pt} {\kern 1pt} {\kern 1pt} {\{v_m\}} \\
	& {\rm s.t.}~{f_{1,k}}\left( {\bf{v}} \right) + {\sigma ^2} \le \left( {{1}/{{{r_{k,{\rm{th}}}}}} + 1} \right){{f}_{2,k}^{{\rm{lb}}}}\left( {\bf{v}} \right),k \in {\cal K},\label{P3_3_const1}\\
&	\qquad  {\rm{tr}}\left( {{\bf{AR}}{{\bf{A}}^H}} \right) \ge {r_{r,{\rm{th}}}}\left( {{{\bf{v}}^T}\left( {\left( {{\bf{G}}_r^H{{\bf{G}}_r}} \right) \odot {{\left( {{{\bf{G}}_t}{\bf{RG}}_t^H} \right)}^T}} \right){{\bf{v}}^*}{\rm{ + }}{{ \sigma }^2}}N_r \right),\label{P3_3_const2}\\
	&\qquad \eqref{P3_phase}.
	\end{align}
\end{subequations}
 It is  observed that all constraints    are convex, problem \eqref{P3_3} is thus convex. However,    problem \eqref{P3_3} has no explicit objective function.  To achieve a better converged solution, we further transform problem  \eqref{P3_3} into an optimization problem with an explicit objective function to obtain a more efficient IRS phase-shift solution. Specifically, by introducing  auxiliary non-negative optimization variables  $\{{\eta _k},k \in {\cal K}\}$ associated with \eqref{P3_3_const1} and ${\eta _0}$ associated with \eqref{P3_3_const2}, problem \eqref{P3_3} can be recast as
  \begin{subequations} \label{P4_3}
	\begin{align}
	&\mathop {\max }\limits_{{\eta _0} \ge 0, {\eta_k} \ge 0,{\bf{v}}} \sum\limits_{k = 1}^K {{\eta _k}}+  {\eta _0} \\
	& {\rm s.t.}~{\eta _k}+{{f}_{1,k}}\left( {\bf{v}} \right) + {\sigma ^2} \le \left( {{1}/{{{r_{k,{\rm{th}}}}}} + 1} \right){f_{2,k}^{{\rm{lb}}}}\left( {\bf{v}} \right),k \in {\cal K},\label{P4_3_const1}\\
	&\qquad {\rm{tr}}\left( {{\bf{AR}}{{\bf{A}}^H}} \right) \ge {r_{r,{\rm{th}}}}\left( {{{\bf{v}}^T}\left( {\left( {{\bf{G}}_r^H{{\bf{G}}_r}} \right) \odot {{\left( {{{\bf{G}}_t}{\bf{RG}}_t^H} \right)}^T}} \right){{\bf{v}}^*}{\rm{ + }}{{ \sigma }^2}N_r} \right) + {\eta _0},\\
	&\qquad \eqref{P3_phase}.
	\end{align}
\end{subequations}
It can be checked that problem \eqref{P4_3} is  convex, which thus can be    solved by convex techniques.
 \begin{algorithm}[!t]
	\caption{SDR-based  algorithm  for solving problem  \eqref{P1}.}	\label{alg2}
	\begin{algorithmic}[1]
		\STATE  \textbf{Initialize} IRS phase shifts   ${{\bf{v}}}$ and  threshold $\varepsilon_1$.
		\STATE  \textbf{repeat}
		\STATE  \quad Update   Transmit covariance matrices   $\left\{ {{{\mathbf{W}}_{c,k}},{{\mathbf{Z}}_{r}}} \right\}$ by solving  problem \eqref{P3_1}.
		\STATE  \quad  Update IRS phase shifts $\{v_m\}$  by solving  problem   \eqref{P4_3}.    
		\STATE \textbf{until}   the fractional decrease of the objective value of problem \eqref{P3} is below $\varepsilon_1$.
		\STATE Reconstruct  phase shift as $v_m^{{\text{opt}}} = \frac{{{v_m}}}{{\left| {{v_m}} \right|}}, m\in {\cal M}$. Then, using this new reconstructed solution   to
		solve  the resulting transmit power minimization problem.
		\STATE Construct  the rank-one solution of communication beamformers  based on \eqref{appendix3_1} and recover radar beamformer based on \eqref{appendix3_2}.
			
	\end{algorithmic}
\end{algorithm}
\subsection{Overall Algorithm}
Based on the above two subproblems, we  alternately optimize each subproblem in an iterative way until convergence is achieved.
It is worth pointing out that the converged solution may not satisfy unit-modulus   constraint as well as rank-one solution of communication beamformers. As such, additional operations are required. To be specific,  we first   normalize the amplitudes of IRS
phase shifts to be   one and  solve  the resulting Radcom BS transmit power minimization problem based on the new  constructed  IRS phase shifts. Then,
we  check   whether the rank of ${{{\mathbf{W}}_{c,k}}}, k\in {\cal K}$ equals to one or not. If the rank of ${{{\mathbf{W}}_{c,k}}}$ is one, then we can obtain the optimal ${{{\mathbf{w}}_{c,k}}}$ by  performing  eigenvalue decomposition on ${{{\mathbf{W}}_{c,k}}}$. For the high rank solution (larger than one) of ${{{\mathbf{W}}_{c,k}}}$,  
the traditional method to extract a rank-one solution from ${{{\mathbf{W}}_{c,k}}}$  is    applying  the Gaussian randomization
technique  \cite{sidiropoulos2006transmit}. However, it inevitably incurs performance loss   as well as  high computational complexity. Fortunately, the following theorem  shows that there always exists the  rank-one solution of ${{{\mathbf{W}}_{c,k}}}, k \in {\cal K}$ to problem \eqref{P3}.

\textbf{\emph{Theorem 2: }}There always exists  a converged communication beamformer solution, denoted as ${{\bf{\hat W}}_{c,k}},k\in{\cal K}$,  satisfying  ${\rm{rank}}\left( {{\bf{\hat W}}_{c,k}^{}} \right)=1,k\in{\cal K}$.

\hspace*{\parindent}\textit{Proof}: Please refer to Appendix~C. \hfill\rule{2.7mm}{2.7mm}

Theorem 2 shows that the Gaussian randomization technique is not needed in general.  We can construct  the rank-one solution of communication beamformer  based on \eqref{appendix3_1} and recover the radar beamformer based on \eqref{appendix3_2}.

The details  are  summarized in Algorithm~\ref{alg2}.  The main computational complexity of Algorithm~\ref{alg2} is given by  ${\cal O}\left( {{L_{{\rm{total}}}}{{\left( {\left( {K{\text{ + }}1} \right)N_t^2} \right)}^{3.5}} + {{\left( {M + K+1} \right)}^{3.5}}} \right)$, where ${{L_{{\rm{total}}}}}$ denotes the total number of iterations required for reaching convergence.
\subsection{Special Case Discussion}
It  still remains unknown whether the radar signals  are really needed in  problem \eqref{P1}. Below, we make an in-depth analysis on this question.

\textbf{\emph{Theorem 3: }}For the special case of problem \eqref{P1} without constraint \eqref{P1_const3}, the  optimal radar beamformer  satisfies $ {{{\mathbf{w}}_{r,i}^{\rm opt}}}={\bf 0}, i\in {{\cal N}_t}$.

\hspace*{\parindent}\textit{Proof}: This result can be directly derived from Theorem 1, and is omitted  for brevity.  \hfill\rule{2.7mm}{2.7mm}

Together with Theorem 1 and Theorem 3, we can conclude that the dedicated  radar signals are not needed regardless of the interference  provided that the cross-correlation design constraint is not considered.

\textbf{\emph{Theorem 4: }} For the SDR of  problem \eqref{P1}, i.e., problem \eqref{P3}, the optimal     radar covariance matrix  satisfies  ${\bf Z}_r^{\rm opt}={\bf 0}$.

\hspace*{\parindent}\textit{Proof}: Please refer to Appendix~D.  \hfill\rule{2.7mm}{2.7mm}

Note that although the SDR of problem \eqref{P1}, i.e., problem  \eqref{P3},  is equivalent to problem \eqref{P3}\textit{-new} defined in Appendix~D, the  reconstructed rank-one approach proposed  in Appendix~C is no longer satisfied for   problem \eqref{P3}\textit{-new}. In general, the SDR tightness for problem \eqref{P3}\textit{-new} may not hold due to the limited  degrees of freedom of the transmitted signals. As  a result, 
the Gaussian randomization technique   may be required to reconstruct the rank-one solution and the performance loss is inevitably incurred \cite{sidiropoulos2006transmit}. This result indicates that the dedicated radar signals may be required provided that the cross-correlation design constraint is  considered.

\section{Numerical Results}
In this section, we provide numerical results    to validate   the performance of our proposed joint design of passive and active beamforming in the Radcom system. A three dimensional   coordinate setup measured in meter (m) is considered, where the Radcom BS is located at $( 0,0,3.5)~\rm m$  and the users are uniformly and randomly distributed in a circle centered at $\left( {d_x,0,1 } \right)~{\rm m}$  with a radius $2~\rm m$, while the IRS is deployed right above the center of the users at $(d_x , 0, 3.5 )~{\rm m}$, where $d_x$   denotes the horizontal location   along $x$-axis.
The distance-dependent path loss model is given by $L\left( {\hat d} \right) = {c_0}{\left( {{\hat d}/{d_0}} \right)^{ - \alpha }}$,
where ${c_0} =-30~{\rm dB}$ is the path loss at the reference distance $d_0=1$ m,  ${\hat d}$ is the link distance, and $\alpha$ is the path loss exponent. We assume that  the Radcom BS-IRS link and  the IRS-user link    follow Rician fading with a Rician factor of $3~{\rm dB}$ and a  path loss exponent of $2.2$, while the  BS-user link  follows Rayleigh  fading with a  path loss exponent of $3.6$ due to the surrounding rich scatters. Both the transmit and receive antennas at the Radcom are uniform linear
arrays with half-wavelength spacing between adjacent antennas, i.e.,  $d=\lambda/2$.  We consider $L=3$ targets which are located at  directions ${\theta _1}{\rm{ = }}{40^ \circ }$, ${\theta _2}{\rm{ = }}{0^ \circ }$, and ${\theta _3}{\rm{ = }}{40^ \circ }$, respectively. In addition, we assume that the communication users have the same SINR constraint, i.e., $r_{c,\rm th}=r_{k,\rm th}, k\in {\cal K}$.
 Unless otherwise specified, we set  $r_{r,\rm th}=10~{\rm dB}$,  $r_{c,\rm th}=20~{\rm dB}$, $N_t=N_r=8$, $d_x=50~{\rm m}$,  ${\sigma ^2}{\rm{ =  - }}80~{\rm dBm}$,  $\sigma _\beta ^2 =  - 70~{\rm dBm}$, $\rho=100$, $c=0.85$, , $\varepsilon_1=10^{-3}$, and $\varepsilon_2=10^{-7}$.
\begin{figure}[!t]
	\centering
	\begin{minipage}[t]{0.45\textwidth}
		\centering
		\includegraphics[width=3.2in]{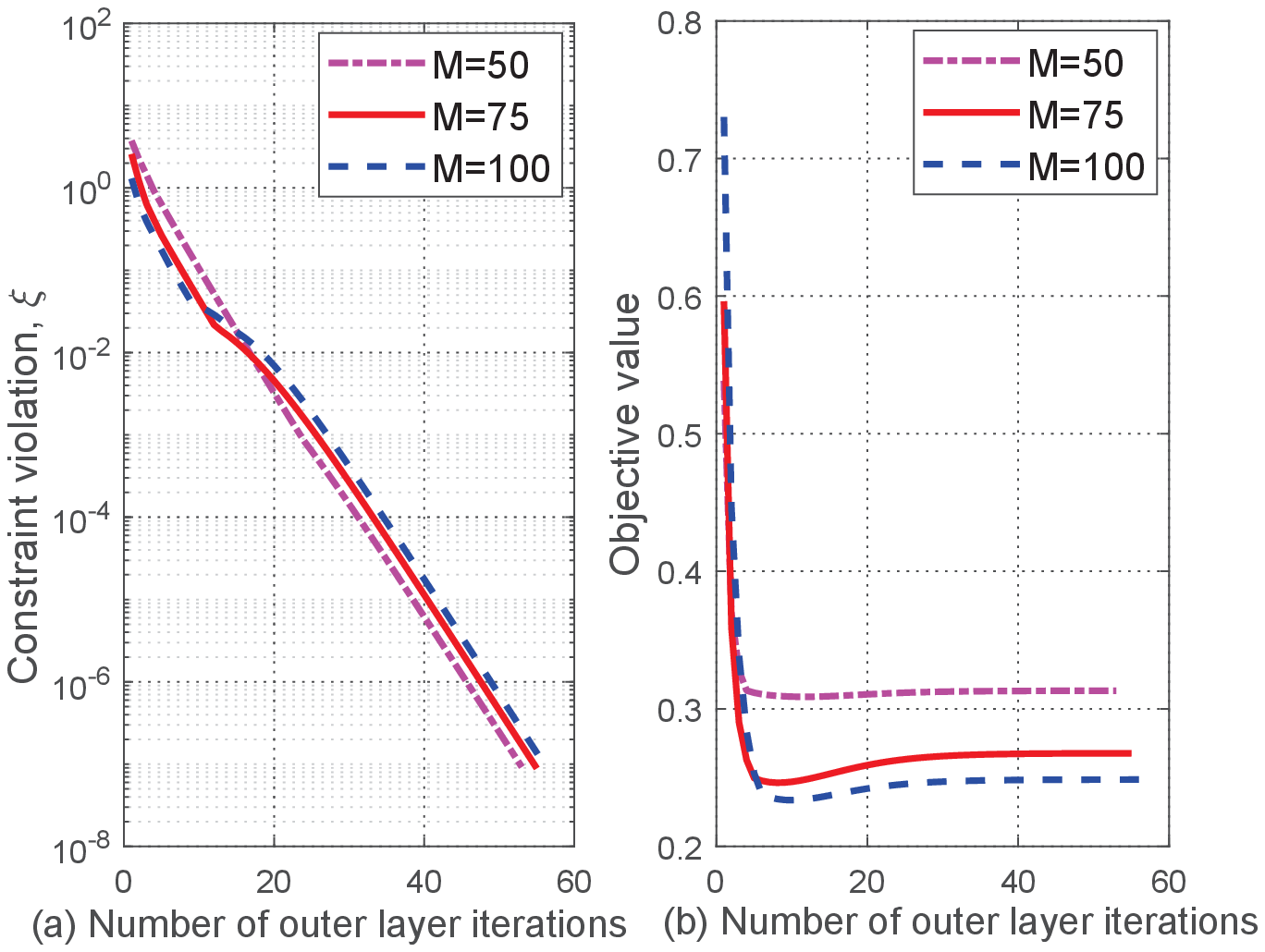}
		\caption{Convergence behaviour of Algorithm~\ref{alg1}.} \label{figconvergencepenlaty}
	\end{minipage}
	\hspace{10pt}
	\begin{minipage}[t]{0.45\textwidth}
		\centering
		\includegraphics[width=3.2in]{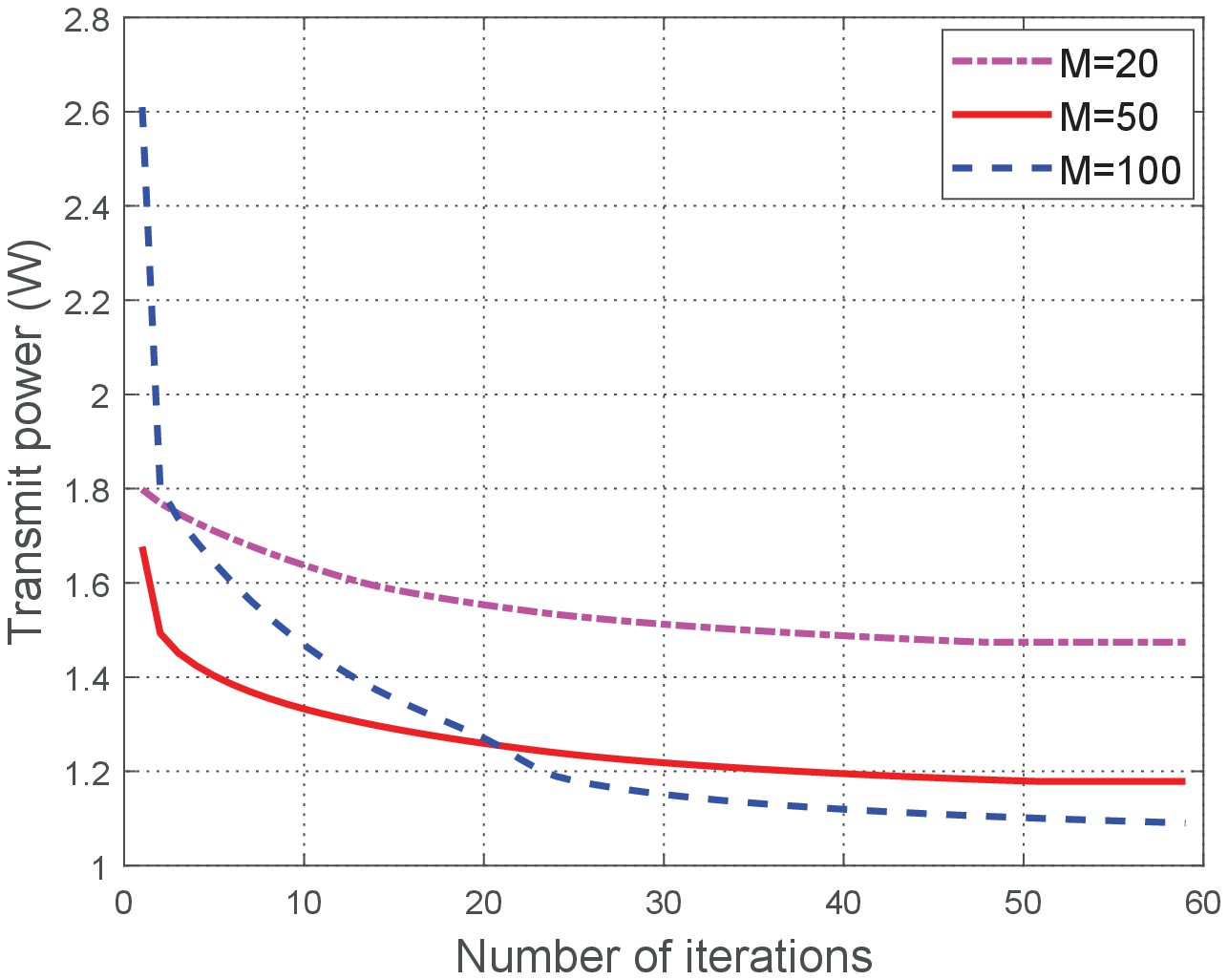}
		\caption{Convergence behaviour of Algorithm~\ref{alg2}.}\label{figconvergenceSDP}
	\end{minipage}
	\vspace{-10pt} 
\end{figure}
\subsection{Convergence Behavior of the Proposed Two Algorithms}
Before discussing the system performance, we first verify the  effectiveness of the proposed penalty-based and SDR-based algorithms, i.e., Algorithm~\ref{alg1} and Algorithm~\ref{alg2}, respectively. 

Fig.~\ref{figconvergencepenlaty} shows the constraint violation and convergence  of Algorithm~\ref{alg1} by solving problem \eqref{P1_without} with $K=5$ for  different number of IRS reflecting elements, namely, $M=50$, $M=75$, and $M=100$. From Fig.~\ref{figconvergencepenlaty}(a), it is observed that constraint violation $\xi$ decreases fast and reaches the predefined accuracy $10^{-7}$ after about $55$ iterations for $M=50$, which indicates that ${\left| {{\bf{h}}_k^H{{\bf{w}}_{c,i}} - x_{k,i}^c} \right|^2}$ and $\left\| {{\bf{A}}{{\bf{w}}_{c,i}} - {\bf{y}}_i^c} \right\|_\infty ^2, i \in {\cal K},k \in {\cal K},$ in \eqref{terminationindicator} are forced to approach zero. As such,
the equality constraint \eqref{P2_const3} in problem \eqref{P2} is eventually satisfied.  Even for large $M$, e.g., $M=100$, the number of
outer layer iterations for reaching the predefined accuracy is about $57$ iterations, which again
demonstrates the  effectiveness  of Algorithm~\ref{alg1}.
In Fig.~\ref{figconvergencepenlaty}(b), we can observe that the objective value \eqref{P2_penalty_obj} is not monotonically decreasing with the number of outer layer  iterations and   fluctuates during the intermediate iterations. This is mainly because when the penalty coefficient  $\rho$ is relatively large,  the  obtained solution  does not satisfy the equality in \eqref{P2_const3}, thus resulting in the oscillatory
behavior. However, as $\rho$ becomes very small, the constraint violation  is forced to approach the  predefined accuracy. Thus, Algorithm~\ref{alg1} is guaranteed to converge finally. 

Fig.~\ref{figconvergenceSDP} shows the convergence of Algorithm~\ref{alg2} by solving problem \eqref{P3}  with  $K=5$ and $\epsilon_{\rm th}=\infty$ (i.e., ignore  cross-correlation constraint \eqref{P3_const3}) for    different  $M$. It is observed that the required transmit power is monotonically decreasing with the number of iterations  and converges about $58$ iterations for different setups. This is expected since each subproblem is optimally/locally solved in each iteration, which  results  in a non-increasing objective value over iterations. In addition, the objective value is lower-bounded   by a finite value due to the minimum  SINR   required by both users  and targets. Thus, Algorithm~\ref{alg2} is guaranteed to converge finally. 
\subsection{IRS-aided Radar and Communication}
In this subsection, we  compare our proposed scheme with several benchmark schemes     under different setups. We adopt the following schemes for comparison: (a) No interference: this corresponds to case I and we study four cases, namely, ``Commun\_only, penalty, no interference'', ``Penalty, no interference'',   ``SDP, no interference'', and ``SDP, no IRS, no interference''. For ``Commun\_only, penalty, no interference'',  only communication signals are  transmitted  by the Radcom BS and the resulting problem  is solved by Algorithm~\ref{alg1}; For ``Penalty, no interference'', both communication signals and radar signals  are transmitted; For ``SDP, no interference'', both communication signals and radar signals  are transmitted and   the resulting problem is solved by Algorithm~\ref{alg2}; For ``SDP, no IRS, no interference'', the IRS is not used.
 (b) Interference: this corresponds to   case  II and we   study two schemes, namely, ``SDP, interference'' and ``Commun\_only, SDP, interference'', and  the corresponding problems are solved by Algorithm~\ref{alg2}. 
 \begin{figure}[!t]
 	\centering
 	\begin{minipage}[t]{0.45\textwidth}
 		\centering
 		\includegraphics[width=3.2in]{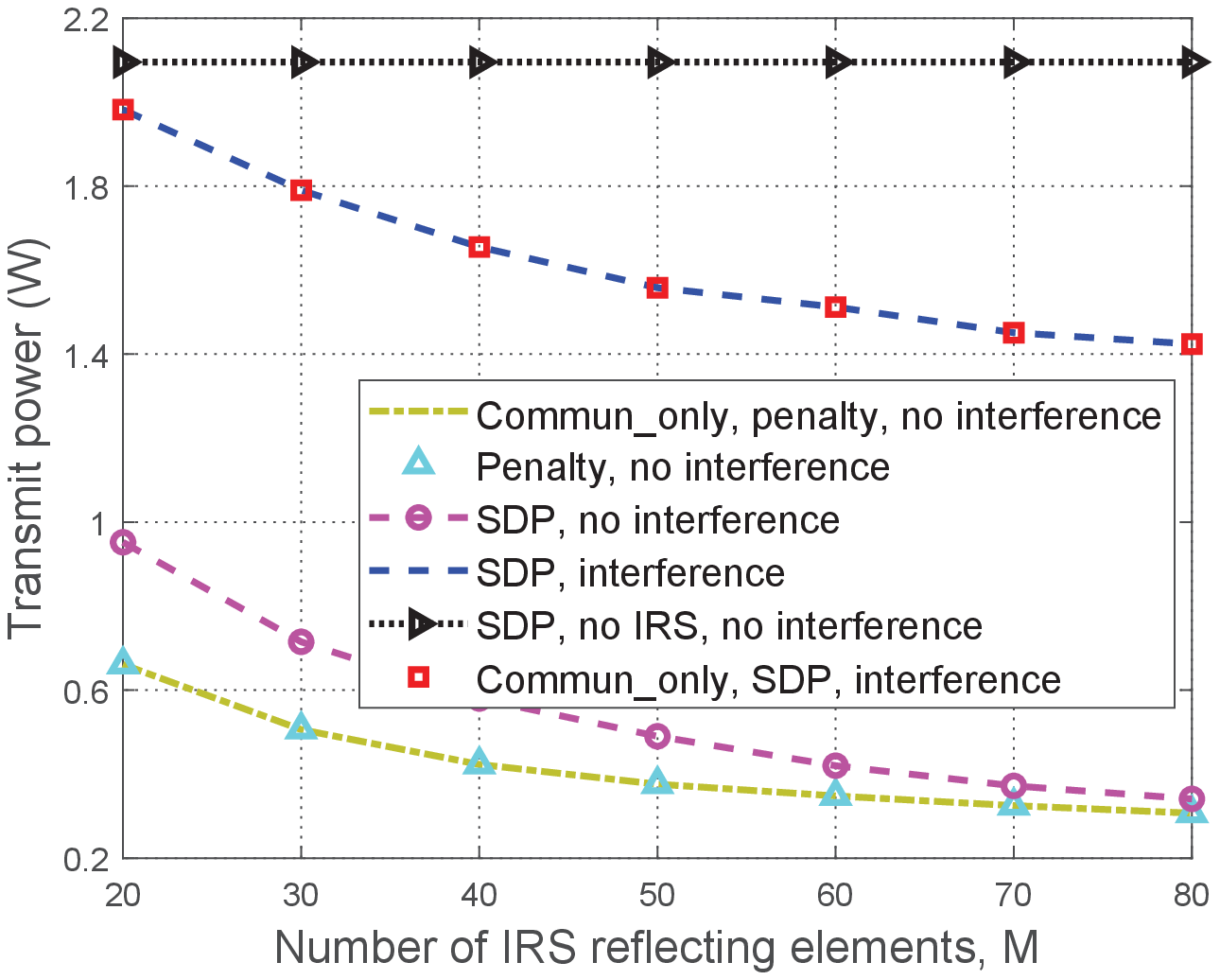}
 		\caption{Transmit power   versus the number of IRS elements.} \label{figvsM}
 	\end{minipage}
 	\hspace{10pt}
 	\begin{minipage}[t]{0.45\textwidth}
 		\centering
 		\includegraphics[width=3.2in]{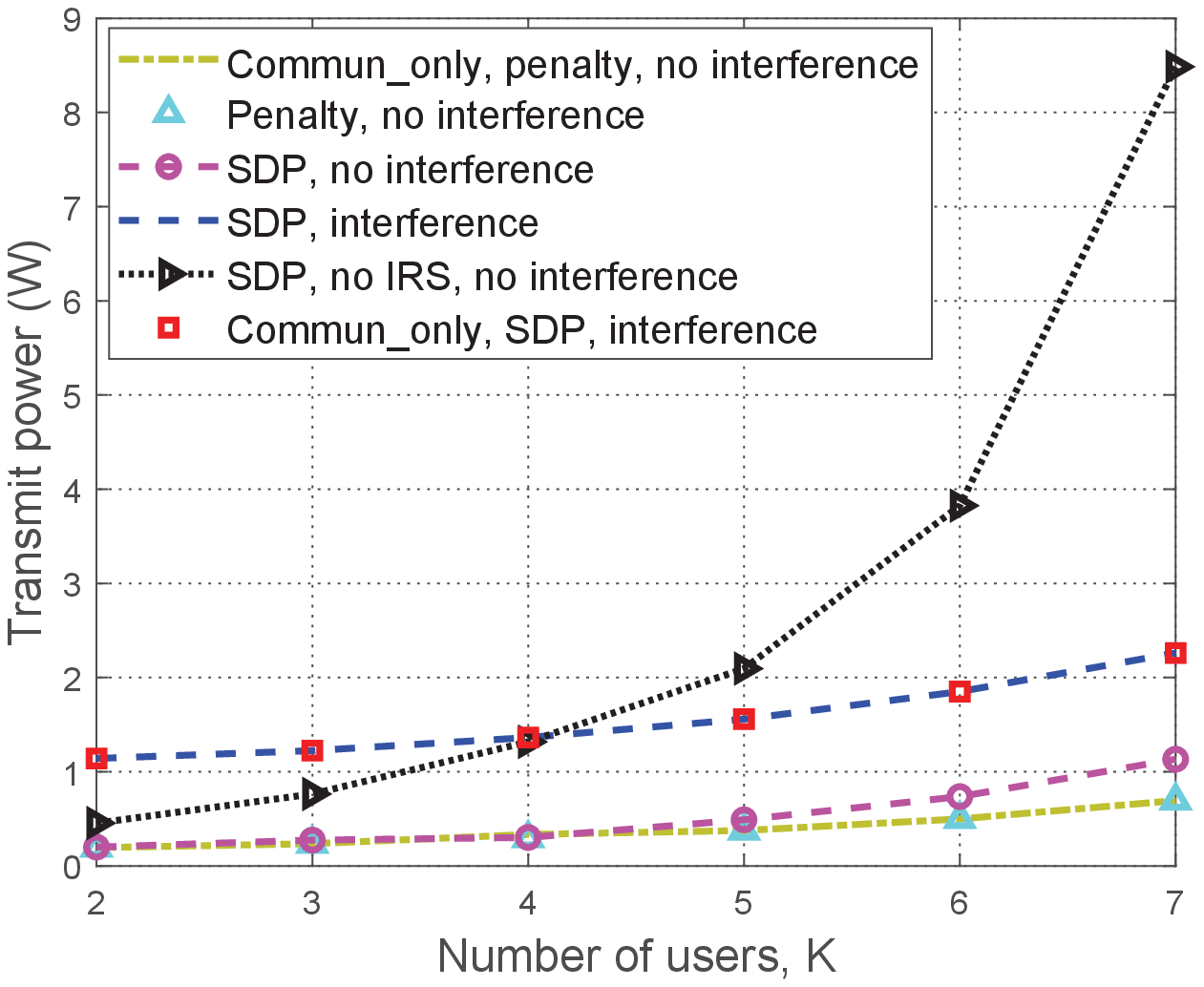}
 		\caption{Transmit power   versus the number of   users.}\label{figvsK}
 	\end{minipage}
 	\vspace{-10pt} 
 \end{figure}
\subsubsection{Effect of Number of IRS Reflecting Elements} In Fig.~\ref{figvsM},  we compare the transmit power obtained by all schemes  versus $M$ with $K=5$ and $\epsilon_{\rm th}=\infty$. First, it is observed that with the optimization of IRS phase shifts, the required transmit power  monotonically decreases with $M$, even when the interference exists. This is because that  installing more reflecting elements at the IRS is able to provide higher passive beamforming gain  towards the desired users, thereby reducing transmit power. Second, it is observed that with the IRS, the schemes without   interference significantly outperform   those with 
interference as expected.  Third, the ``Commun\_only, penalty, no interference'' scheme  achieves the same performance with the  ``Penalty, no interference'' scheme, which implies the  radar signals are unnecessary and justifies Theorem 1. In addition, we   also observe  the same results  for the case  with interference, which justifies Theorem 3. Last, the penalty-based algorithm achieves lower transmit power than the SDP-based algorithm. This is because that with the proper variables partitioning, there is no constraint coupling between the variables in different blocks as shown in Algorithm~\ref{alg1},  while  it does not hold    in Algorithm~\ref{alg2}.  
\subsubsection{Effect of Number of Users} In Fig.~\ref{figvsK},  we compare the transmit power obtained by all schemes  versus $K$ with $M=50$ and $\epsilon_{\rm th}=\infty$.  It is observed that the required transmit power obtained by all schemes  is monotonically increasing  as $K$ increases.   This is because that the required transmit power highly depends on the  user  who has the worst channel quality
to satisfy the minimum  SINR. In addition, we observe that the scheme without IRS requires much higher transmit power than those with   IRS in the  without interference case, especially when $K$ becomes large,  which demonstrates the benefits of applying IRS in the Radcom system.
 Furthermore, we   observe that   the scheme with only  communication signal transmission  achieves the same transmit power with the joint signal transmission, which again justifies Theorem 1 and Theorem 3.
\begin{figure}[!t]
	\centering
	\begin{minipage}[t]{0.45\textwidth}
		\centering
		\includegraphics[width=3.2in]{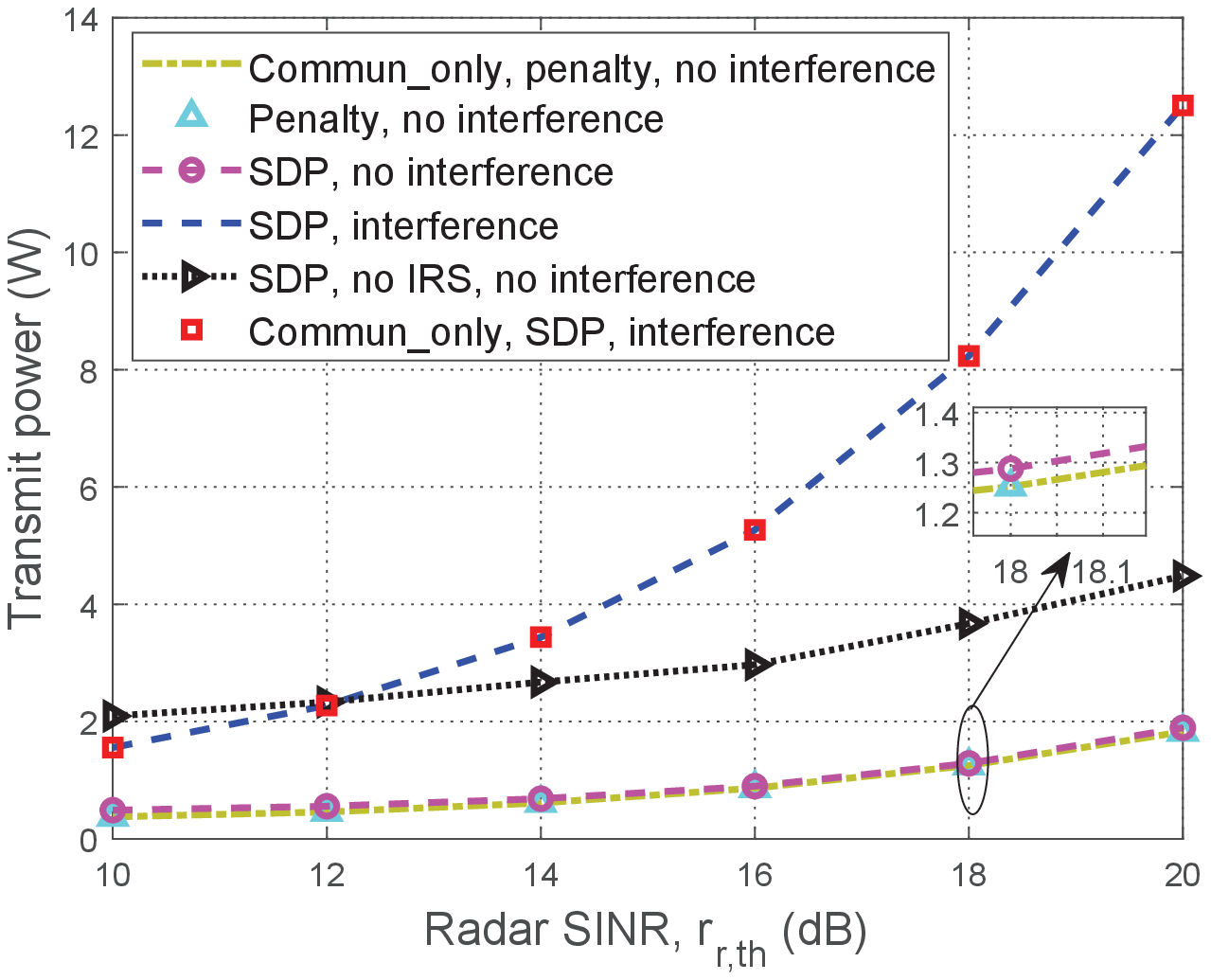}
		\caption{Transmit power   versus   the  Radar SINR.} \label{figvsRadarSNR}
	\end{minipage}
	\hspace{10pt}
	\begin{minipage}[t]{0.45\textwidth}
		\centering
		\includegraphics[width=3.2in]{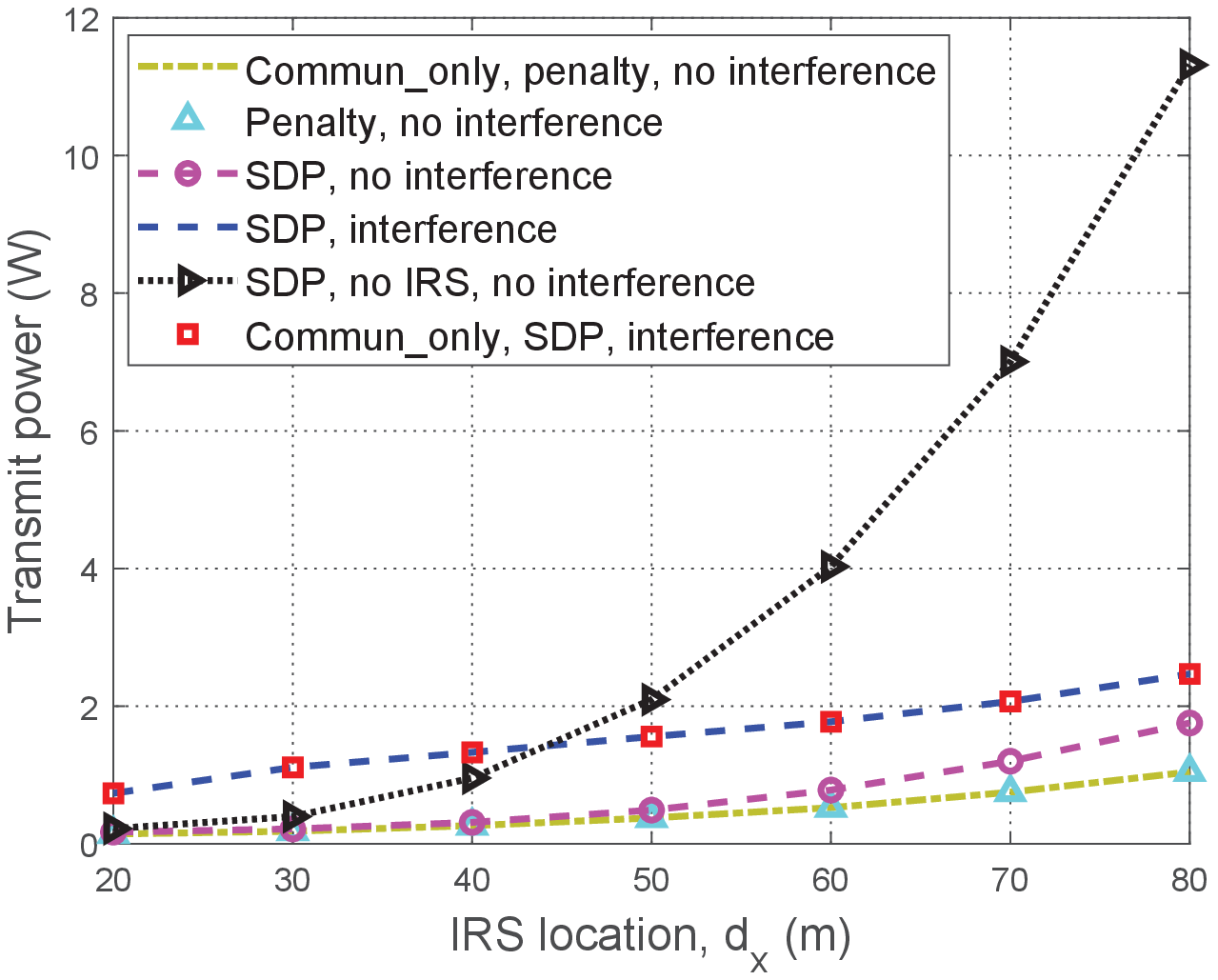}
		\caption{Transmit power   versus  the  Radar SINR.}\label{figvsIRSlocation}
	\end{minipage}
	\vspace{-10pt} 
\end{figure}
\subsubsection{Effect of Radar  SINR} In Fig.~\ref{figvsRadarSNR}, we study the impact of   radar  SINR  $r_{r,\rm th}$ on the  transmit power  required at the Radcom BS with $K=5$, $M=50$, $\epsilon_{\rm th}=\infty$, and $r_{c,\rm th}=20~{\rm dB}$.  We   observe that the required transmit power  obtained by   the scheme  with interference remarkably increases as compared with that obtained by the scheme  without interference, especially when   the required $r_{r,\rm th}$ is high. This is because that as $r_{r,\rm th}$ becomes large,  the interference introduced by the IRS will be more prominent, which thus requires higher transmit power to satisfy the radar SINR.
 However, if the interference can be perfectly canceled, the scheme with the IRS  achieves much  lower transmit power than that  without IRS due to the high passive beamforming gains brought by the IRS.

\subsubsection{Effect of IRS deployment}In Fig.~\ref{figvsIRSlocation}, we study the impact of IRS deployment/location $d_x$ on the system performance with $M=50$, $K=5$, and $\epsilon_{\rm th}=\infty$. We observe that as  $d_x$ increases, i.e., the distance between the IRS and the Radcom BS becomes larger, the required transmit power is remarkably increased by the scheme  without IRS due to the high path-loss attenuation. In addition,  the performance gap between ``SDP, no interference'' and ``SDP, no IRS, no interference'' becomes more pronounced when the IRS is far away from the Radcom BS, which   further demonstrates the benefits brought by the IRS. However, this result does not hold for the schemes  with interference. To be specific, when $d_x\le45$, the  ``SDP, interference'' scheme  consumes more transmit power than the  `SDP, no IRS, no interference'' scheme, while when $d_x\ge45$, the ``SDP, interference'' scheme  saves  more transmit power than the `SDP, no IRS, no interference'' scheme.  This is because as the IRS is deployed close to the Radcom, i.e., $d_x\le45$,  the interference introduced by the IRS is significant, thus impairing the system performance. However, as the IRS  is far away from the Radcom, i.e., $d_x\ge45$, the    interference introduced by the IRS becomes small. To  see  it  clearly, it is observed  that the performance gap between  ``SDP, interference'' and ``SDP, no interference'' becomes smaller as $d_x$ increases, which indicates that the impact of  interference introduced by the IRS on the Radcom becomes smaller. 

\begin{figure}[!t]
	\centering
	\begin{minipage}[t]{0.45\textwidth}
		\centering
		\includegraphics[width=3.6in]{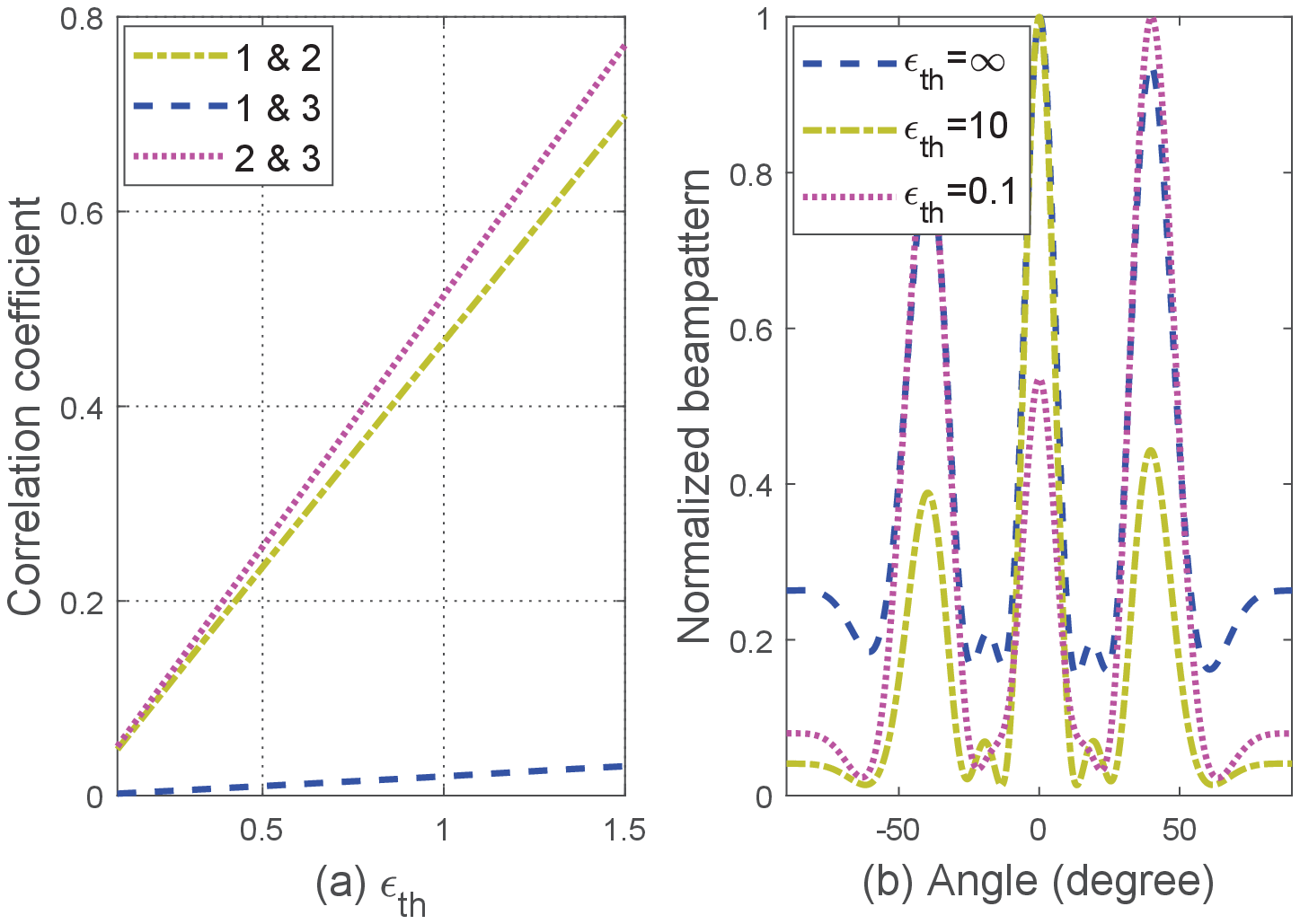}
		\caption{Beampattern   design.} \label{figvscorrelationefficient}
	\end{minipage}
	\hspace{25pt}
	\begin{minipage}[t]{0.45\textwidth}
		\centering
		\includegraphics[width=3.2in]{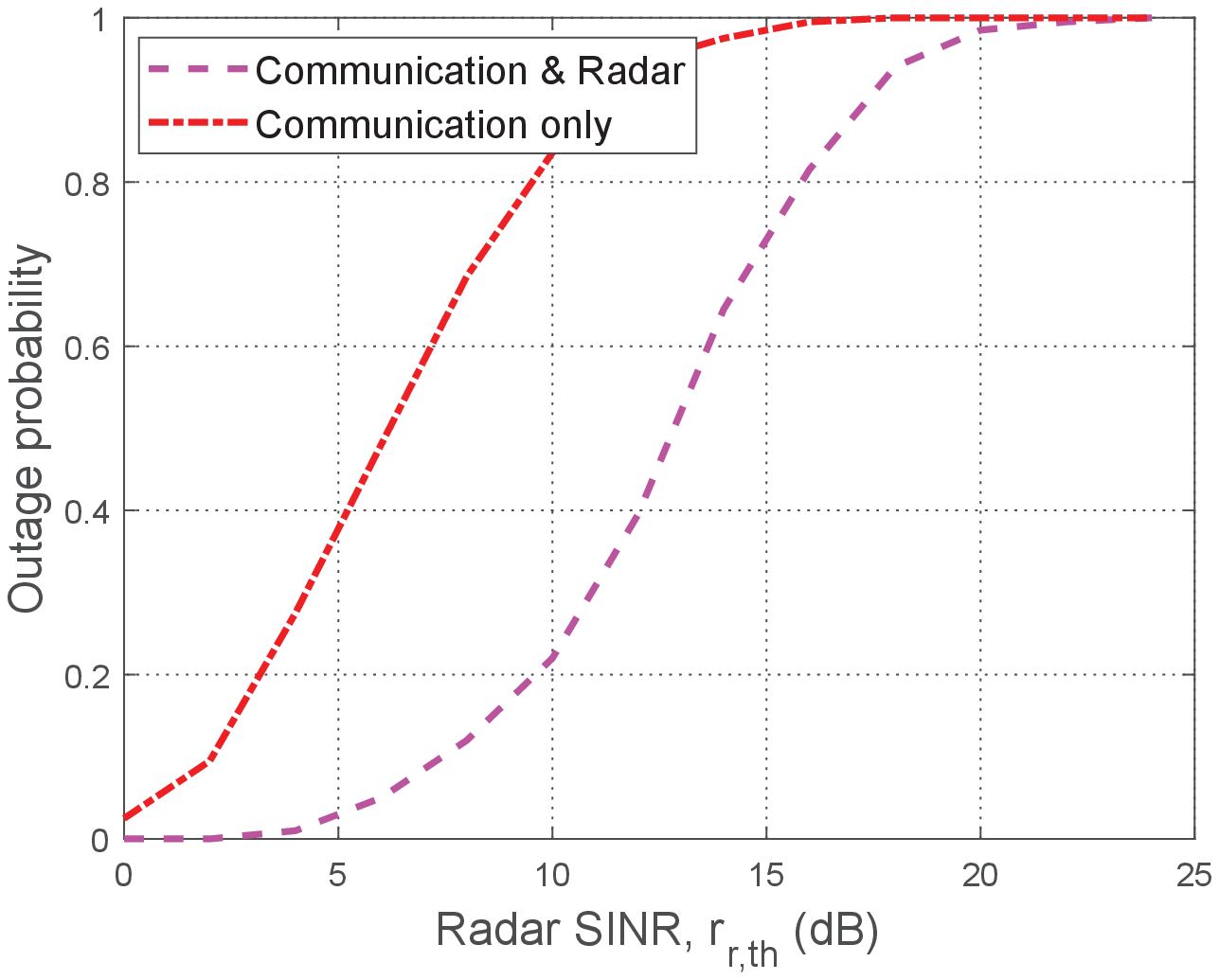}
		\caption{Outage probability versus Radar SINR.}\label{figvsoutage}
	\end{minipage}
	\vspace{-10pt} 
\end{figure}
\subsubsection{Effect of Cross-Correlation Constraint} In Fig.~\ref{figvscorrelationefficient}(a), we study the impact of $\epsilon_{\rm th}$ on the cross-correlation coefficients of the three target reflected signals. It is observed that when $\epsilon_{\rm th}$ approaches zero, i.e., $\xi_{\rm th}=0.1$, which implies that a stringent cross-correlation is imposed,  and  all cross-correlation coefficients are very small. However, as $\epsilon_{\rm th}$  becomes large, the first and second reflected signals, i.e., $1~\&~2$, and the second and third reflected signals, i.e., $2~\&~3$, are highly correlated, which can degrade significantly the performance of any adaptive technique for the multi-target radar detection \cite{liu2006lu}. 
An example of  the normalized magnitudes of beampatterns obtained for different $\epsilon_{\rm th}$ under  ${\beta _l} = {\sigma _\beta }, l\in {\cal L}$ is studied in Fig.~~\ref{figvscorrelationefficient}(b). We can observe that all the schemes with different $\epsilon_{\rm th}$ can track targets well, and the 
beampatterns obtained with  $\epsilon_{\rm th}=\infty$ is better than that with the other cross-correlation constraint, i.e.,  $\epsilon_{\rm th}=0.1$ and $\epsilon_{\rm th}=10$.
\subsubsection{Single Waveform versus Joint Waveforms} To evaluate  the impact of the single weaveform design and the joint waveform design on the system performance. In Fig.~\ref{figvsoutage}, we study the outage probability versus  Radar SINR $r_{r,\rm th}$ in case II for different transmit beamforming  schemes with $K=5$, $M=50$, $d_x=20$,  and $\epsilon_{\rm th}=1$.  Two schemes are  compared: 1) Communication \& Radar: both the communication and radar signals are used for transmission and the resulting problem is solved by Algorithm~\ref{alg2}; 2) Communication only: only the communication signal is used for transmission and the resulting problem is solved by Algorithm~\ref{alg2} but the Gaussian randomization technique is applied for reconstructing rank-one solution (Here, $1000$ Gaussian randomization realizations are performed). It is observed that the outage probability obtained by  the two schemes increases as $r_{r,\rm th}$ increases  and finally approaches $1$ for large $r_{r,\rm th}$. This is expected since the higher transmit power is required for satisfying the stringent Radar SINR constraint, thereby potentially increasing the cross-correlation  coefficient  in \eqref{P3_const3} and making the problem infeasible with a higher probability. In addition, 
it is  observed that the ``Communication \& Radar'' scheme  performs better than the ``Communication only'' scheme,  which indicates that  the radar signal  is  useful for system design. 
\section{Conclusion}
In this paper, we   studied the joint design of  active beamforming  and passive beamforming for an IRS-aided Radcom system. The transmit power minimization problems for two cases, i.e., case I and case II,  based on  the presence or absence of  the radar cross-correlation  design   and   the  interference introduced by the IRS   on  the Radcom BS were formulated.
We first studied  case I and  proved that the dedicated radar signals are not required, and then   proposed     a  penalty-based algorithm to solve the formulated non-convex optimization problem. Then, we studied case II and showed that the dedicated radar signals are  required in general to enhance the system performance, and  an SDR-based AO algorithm is proposed to solve this challenging optimization problem. Simulation results demonstrated the benefits of the IRS used for enhancing the performance of  the Radcom system.  In addition, the results also showed for case II that adopting  dedicated radar signals  at the Radcom BS can significantly reduce the system outage probability as compared to the case without adopting the dedicated radar signals.

\section*{Appendix A: \textsc{Proof of theorem 1}}\label{appendix0}
We prove Theorem~1 by solving the SDR of problem \eqref{P1_without}. Specifically, 
define ${{\mathbf{Z}}_r} = {{\mathbf{W}}_r}{\mathbf{W}}_r^H$ and    ${{\bf{W}}_{c,k}} = {{\bf{w}}_{c,k}}{\bf{w}}_{c,k}^H,k \in {\cal K}$, which need to satisfy  ${\bf Z}_r\succeq{\bf 0}, $${{\bf{W}}_{c,k}} \succeq {\bf{0}}$, and ${\rm{rank}}\left( {{{\bf{W}}_{c,k}}} \right) = 1$.
By ignoring the above rank-one constraint on ${{\mathbf{W}}_{c,k}}$'s, the SDR of problem \eqref{P1_without} for any given IRS phase shifts $\bf v$
is given by 
\begin{subequations} \label{P1_without_SDR}
	\begin{align}
	&\mathop {\min }\limits_{\left\{ {{{\mathbf{W}}_{c,k}}} \right\},{{\mathbf{Z}}_r}} \sum\limits_{k = 1}^K {{\text{tr}}} \left( {{{\mathbf{W}}_{c,k}}} \right) + {\text{tr}}\left( {{{\mathbf{Z}}_r}} \right) \\
	& {\rm s.t.}~{\mathbf{h}}_k^H\left( {\sum\limits_{k = 1}^K {{{\mathbf{W}}_{c,k}}}  + {{\mathbf{Z}}_r}} \right){{\mathbf{h}}_k} + {\sigma ^2} \le  \left( {\frac{1}{{{r_{k,{\text{th}}}}}} + 1} \right){\mathbf{h}}_k^H{{\mathbf{W}}_{c,k}}{{\mathbf{h}}_k},k \in {\cal K},\label{P1_without_SDRconst1}\\
	&\qquad {\text{tr}}\left( {{\mathbf{A}}\left( {\sum\limits_{k = 1}^K {{{\mathbf{W}}_{c,k}}}  + {{\mathbf{Z}}_r}} \right){{\mathbf{A}}^H}} \right)\ge {r_{r,{\rm{th}}}}{{ \sigma }^2}.\label{P1_without_SDR_const2}
	\end{align}
\end{subequations}
It is not difficult to see that problem \eqref{P1_without_SDR} is a  semidefinite programming (SDP) problem and  satisfies the Slater's condition, which indicates that the  duality gap is zero. Thus, we consider the following Lagrangian of problem \eqref{P1_without_SDR} given by 
\begin{align}
{{\cal L}_1}\left\{ {\left\{ {{{\mathbf{W}}_{c,k}}} \right\},{{\mathbf{Z}}_r},\left\{ {{\lambda _{1,k}}} \right\},\mu } \right\} = \sum\limits_{k = 1}^K {{\text{tr}}} \left( {{{\mathbf{B}}_k}{{\mathbf{W}}_{c,k}}} \right) + {\text{tr}}\left( {{\mathbf{C}}{{\mathbf{Z}}_r}} \right) + \sum\limits_{k = 1}^K {{\lambda _{1,k}}{\sigma ^2}}  + \mu  \sigma^2{r_{r,{\text{th}}}}, \label{P1_without_SDR_lag}
\end{align}
where 
\begin{align}
&{{\mathbf{B}}_k} = {{\mathbf{I}}_{{N_t}}} + \sum\limits_{i \ne k}^K {{\lambda _{1,i}}{{\mathbf{h}}_i}{\mathbf{h}}_i^H}  - {{{\lambda _{1,k}}{{\mathbf{h}}_k}{\mathbf{h}}_k^H}}/{{{r_{k,{\text{th}}}}}} - \mu {{\mathbf{A}}^H}{\mathbf{A}},k \in {\cal K}, \label{appendix0_3}\\
&{\mathbf{C}} = {{\mathbf{I}}_{{N_t}}} + \sum\limits_{k = 1}^K {{\lambda _{1,k}}{{\mathbf{h}}_k}{\mathbf{h}}_k^H}  - \mu {{\mathbf{A}}^H}{\mathbf{A}},\label{appendix0_4}
\end{align}
and ${\left\{ {{\lambda _{1,k}}} \ge 0\right\}}$ and $\mu \ge 0 $ are the dual variables associated with  constraints  \eqref{P1_without_SDRconst1} and \eqref{P1_without_SDR_const2}, respectively. Denote by  dual function  ${f_1}\left\{ {\left\{ {{\lambda _{1,k}}} \right\},\mu } \right\} = \mathop {\min }\limits_{\left\{ {{{\mathbf{W}}_{c,k}}} \right\},{{\mathbf{Z}}_r}} {{\cal L}_1}\left\{ {\left\{ {{{\mathbf{W}}_{c,k}}} \right\},{{\mathbf{Z}}_r},\left\{ {{\lambda _{1,k}}} \right\},\mu } \right\}$, we have the following lemma:

\textbf{\emph{Lemma 1:}}
	To make dual function ${f_1}\left\{ {\left\{ {{\lambda _{1,k}}} \right\},\mu } \right\}$ bounded, we must have 
\begin{align}
{\mathbf{C}} \succeq {\mathbf{0}},~~{{\mathbf{B}}_k} \succeq {\mathbf{0}},k \in {\cal K}.
\end{align}

\hspace*{\parindent}\textit{Proof}: This can be proved by contradiction. Suppose that ${{\mathbf{B}}_k}$ $({\mathbf{C}})$ has at least one negative eigenvalue, we can always  construct a solution of ${{{\mathbf{W}}_{c,k}}}$ $({{{\mathbf{Z}}_r}})$ that has the same eigenvectors  with  ${{\mathbf{B}}_k}$ $({\mathbf{C}})$, while  the  eigenvalues of ${{{\mathbf{W}}_{c,k}}}$ $({{{\mathbf{Z}}_r}})$ corresponding to ${{\mathbf{B}}_k}$ $({\mathbf{C}})$ with  negative eigenvalues  are set to  be positive infinity, resulting in  ${\text{tr}}\left( {{{\mathbf{B}}_k}{{\mathbf{W}}_{c,k}}} \right) \to  - \infty $ and ${\text{tr}}\left( {{\mathbf{C}}{{\mathbf{Z}}_r}} \right) \to  - \infty $. This thus completes the proof. \hfill\rule{2.7mm}{2.7mm}
%
%
%

Accordingly, the dual problem of \eqref{P1_without_SDR} is given by 
\begin{subequations}\label{P1_without_dualproblem}
\begin{align}
	&\mathop {\max }\limits_{\left\{ {{\lambda _{1,k}} \ge 0} \right\},\mu  \ge 0} \sum\limits_{k = 1}^K {{\lambda _{1,k}}{\sigma ^2}}  + \mu  \sigma _r^2{r_{r,{\text{th}}}}\\
	&{\rm s.t.}~{\bf C}\succeq {\mathbf{0}},~~{{\mathbf{B}}_k} \succeq {\mathbf{0}}, k \in {\cal K}.
\end{align}
\end{subequations}
Based on Lemma~$1$,  it is not difficult to prove that at  optimal solutions $\left\{ {{\mathbf{W}}_{c,k}^{{\text{opt}}},{\mathbf{Z}}_r^{{\text{opt}}}} \right\}$ to minimize \eqref{P1_without_SDR_lag} for  fixed dual variables,  the following  equalities must hold:
\begin{align}
{\rm{tr}}\left( {{{\bf{C}}^{{\rm{opt}}}}{\bf{Z}}_r^{{\rm{opt}}}} \right) = 0,\;\;{\rm{tr}}\left( {{\bf{B}}_k^{{\rm{opt}}}{\bf{W}}_{c,k}^{{\rm{opt}}}} \right) = 0,k \in {\cal K},  \label{appendix0_1}
\end{align}
which are equivalent to ${{{\mathbf{C}}^{{\text{opt}}}}{\mathbf{Z}}_r^{{\text{opt}}}}={\bf 0}$ and ${{\mathbf{B}}_k^{{\text{opt}}}{\mathbf{W}}_{c,k}^{{\text{opt}}}}={\bf 0},k \in {\cal K}$.


To prove Theorem~1, we need to prove that the optimal solutions of problem \eqref{P1_without_SDR} should satisfy ${\text{rank}}\left( {{\mathbf{W}}_{c,k}^{{\text{opt}}}} \right) = 1,k \in {\cal K}$, and ${\mathbf{Z}}_r^{{\text{opt}}} = {\mathbf{0}}$. To  proceed it, we 
 consider the following two cases: 1) $\lambda _{1,k}^{{\text{opt}}} = 0$ for   $k \in {\cal K}$; 2) at least one $\lambda _{1,k}^{{\text{opt}}}$ for   $k \in {\cal K}$ is not equal to zero.

For the first case,  it is not difficult to see that ${{\mathbf{C}}^{{\text{opt}}}}{\text{ = }}{\mathbf{B}}_k^{{\text{opt}}}{\text{ = }}{{\mathbf{I}}_{{N_t}}} - {\mu ^{{\text{opt}}}}{{\mathbf{A}}^H}{\mathbf{A}}, k \in {\cal K}$. To guarantee ${{\mathbf{B}}_k^{\rm opt}} \succeq {\mathbf{0}}, k \in {\cal K}$, and maximize dual problem \eqref{P1_without_dualproblem}, the optimal dual  variable ${\mu ^{{\text{opt}}}}$ should satisfy ${\mu ^{{\text{opt}}}} = 1/{\pi _{\max }}$, where ${\pi _{\max }}$ represents the  maximum eigenvalue of ${{\mathbf{A}}^H}{\mathbf{A}}$. Under the assumption that amplitudes  of targets are independently distributed, i.e.,  the non-zero singular values of ${\bf A}^H{\bf A}$ are not the same, ${{{\mathbf{C}}^{{\text{opt}}}}}$ and ${{\mathbf{B}}_k^{{\text{opt}}}}$ must have only one zero eigenvalue  and satisfy 
 ${\text{rank}}\left( {{{\mathbf{C}}^{{\text{opt}}}}} \right) = {\text{rank}}\left( {{\mathbf{B}}_k^{{\text{opt}}}} \right) = {N_t} - 1, k\in {\cal K}$. Denote by ${{\bf v} _{\max }}$ the eigenvector corresponding to the maximum eigenvalue of ${{\mathbf{A}}^H}{\mathbf{A}}$.  It is readily to see that ${\mathbf{Z}}_r^{{\text{opt}}}$ and ${\mathbf{W}}_{c,k}^{{\text{opt}}}$ should all lie in the subspace spanned by ${{\bf v} _{\max }}$. This indicates that  all communication beams  should point towards the targets rather than the communication users and the minimum user SINR requirements  in \eqref{P1_without_SDRconst1} will not be satisfied any more.  Obviously, the case of  $\lambda _{1,k}^{{\text{opt}}} = 0,k \in {\cal K}$, cannot occur here.
 
 For the second case, with  any given ${\mathbf{Z}}_r^{{\text{opt}}} \succeq {\bf 0}$ satisfying \eqref{appendix0_1}, we have 
 \begin{align}
  - \left( {{1}/{{{r_{k,{\text{th}}}}}} + 1} \right){\lambda _{1,k}^{\rm opt}}{\mathbf{h}}_k^H{\mathbf{Z}}_r^{{\text{opt}}}{{\mathbf{h}}_k}& = {\text{tr}}\left( {{{\mathbf{C}}^{{\text{opt}}}}{\mathbf{Z}}_r^{{\text{opt}}} - \left( {{1}/{{{r_{k,{\text{th}}}}}} + 1} \right){\lambda _{1,k}}{{\mathbf{h}}_k}{\mathbf{h}}_k^H{\mathbf{Z}}_r^{{\text{opt}}}} \right)\notag\\
 &= {\text{tr}}\left( {{\mathbf{B}}_k^{{\text{opt}}}{\mathbf{Z}}_r^{{\text{opt}}}} \right) \ge  0,\label{appendix0_2}
 \end{align}
where the first equality follows from \eqref{appendix0_1}, the second equality follows from \eqref{appendix0_3} and \eqref{appendix0_4}, and
 the last inequality holds since both ${{\mathbf{B}}_k^{{\text{opt}}}}$ and ${{\mathbf{Z}}_r^{{\text{opt}}}}$ are positive semidefinite matrices. 
 From \eqref{appendix0_2}, we can derive  $\lambda _{1,k}^{{\text{opt}}}{\text{tr}}\left( {{{\mathbf{h}}_k}{\mathbf{h}}_k^H{\mathbf{Z}}_r^{{\text{opt}}}} \right) = 0$ since $\lambda _{1,k}^{{\text{opt}}} \ge 0$ and ${\mathbf{h}}_k^H{\mathbf{Z}}_r^{{\text{opt}}}{{\mathbf{h}}_k} \ge 0$, $k \in {\cal K}$. As a result, based on  \eqref{appendix0_1} and together with $\lambda _{1,k}^{{\text{opt}}}{\text{tr}}\left( {{{\mathbf{h}}_k}{\mathbf{h}}_k^H{\mathbf{Z}}_r^{{\text{opt}}}} \right) = 0$, we have 
\begin{align}
\left( {{{\mathbf{I}}_{{N_t}}} - {\mu ^{{\text{opt}}}}{{\mathbf{A}}^H}{\mathbf{A}}} \right){\mathbf{Z}}_r^{{\text{opt}}} = \left( {{{\mathbf{I}}_{{N_t}}} + \sum\limits_{k = 1}^K {{\lambda _{1,k}^{\rm opt}}{{\mathbf{h}}_k}{\mathbf{h}}_k^H}  - \mu {{\mathbf{A}}^H}{\mathbf{A}}} \right){\mathbf{Z}}_r^{{\text{opt}}} = {\mathbf{0}}.
\end{align}
Suppose that $\lambda _{1,m}^{{\text{opt}}}$ is the non-zero eigenvalue, i.e., $\lambda _{1,m}^{{\text{opt}}}>0$, it follows that ${{{\mathbf{h}}_m}{\mathbf{h}}_m^H{\mathbf{Z}}_r^{{\text{opt}}}}={\bf 0}$. Since the non-zero singular values of ${\bf A}^H{\bf A}$ are not the same, we have ${\text{rank}}\left( {{{\mathbf{I}}_{{N_t}}} - {\mu ^{{\text{opt}}}}{{\mathbf{A}}^H}{\mathbf{A}}} \right) \ge {N_t} - 1$. This indicates that two matrices ${{{\mathbf{I}}_{{N_t}}} - {\mu ^{{\text{opt}}}}{{\mathbf{A}}^H}{\mathbf{A}}}$ and ${{{\mathbf{h}}_m}{\mathbf{h}}_m^H}$ span the entire space with probability one under the assumption that amplitudes  of targets and user channels are uncorrelated.  As such, we must have ${\mathbf{Z}}_r^{{\text{opt}}} = {\mathbf{0}}$.

Next, we show     to  prove ${\text{rank}}\left( {{\mathbf{W}}_{c,k}^{{\text{opt}}}} \right) = 1,k \in {\cal K}$. On the one hand, based on \eqref{appendix0_1}, it follows that ${\text{rank}}\left( {{\mathbf{B}}_k^{{\text{opt}}}} \right) \le {N_t} - 1$ since ${{\mathbf{W}}_{c,k}^{\rm opt}} \ne {\mathbf{0}}$ (otherwise the communication user SINR constraint \eqref{P1_without_SDRconst1} will not be satisfied).  On the other hand, recall that any  ${\mathbf{Z}}_r^{{\text{opt}}}\succeq {\bf 0}$ satisfying \eqref{appendix0_1} should be $0$, it follows that ${\text{rank}}\left( {{{\mathbf{C}}^{{\text{opt}}}}} \right) = {N_t}$. Based on \eqref{appendix0_3}  and \eqref{appendix0_4}, we have 
\begin{align}
{\text{rank}}\left( {{\mathbf{B}}_k^{{\text{opt}}}} \right)& = {\text{rank}}\left( {{{\mathbf{C}}^{{\text{opt}}}} - \left( {{1}/{{{r_{k,{\text{th}}}}}} + 1} \right)\lambda _{1,k}^{{\text{opt}}}{{\mathbf{h}}_k}{\mathbf{h}}_k^H} \right)\notag\\
&  \ge {\text{rank}}\left( {{{\mathbf{C}}^{{\text{opt}}}}} \right) - {\text{rank}}\left( {\left( {{1}/{{{r_{k,{\text{th}}}}}} + 1} \right)\lambda _{1,k}^{{\text{opt}}}{{\mathbf{h}}_k}{\mathbf{h}}_k^H} \right)\notag\\
& = {N_t} - 1.
\end{align}
Thus, combining arguments ${\mathbf{B}}_k^{{\text{opt}}} \le  {N_t} - 1$ and ${\mathbf{B}}_k^{{\text{opt}}} \ge {N_t} - 1$,  we have ${\mathbf{B}}_k^{{\text{opt}}} = {N_t} - 1,k\in {\cal K}$. Based on \eqref{appendix0_1}, it follows ${\text{rank}}\left( {{\mathbf{W}}_{c,k}^{{\text{opt}}}} \right) = 1,k\in {\cal K}$. Together with the facts that ${\mathbf{Z}}_r^{{\text{opt}}} = {\mathbf{0}}$ and  ${\text{rank}}\left( {{\mathbf{W}}_{c,k}^{{\text{opt}}}} \right) = 1,k\in {\cal K}$, we can conclude that problem \eqref{P1_without_SDR} is equivalent to problem \eqref{P1_without} and no radar beams are required, which completes the proof.

\section*{Appendix B: \textsc{Proof of lemma 2}}\label{appendix1}
To show Lemma~$2$, we expand Lagrangian function \eqref{Lagrangian}   as 
\begin{align}
 {{\cal L}_2}\left( {x_{k,i}^c,{\lambda _{2,k}}} \right) &= \left( {1 - {\lambda _{2,k}}} \right){\left| {x_{k,k}^c} \right|^2} - 2{\text{Re}}\left\{ {x_{k,k}^{c,H}{\mathbf{h}}_k^H{{\mathbf{w}}_{c,k}}} \right\}+{\left| {{\mathbf{h}}_k^H{{\mathbf{w}}_{c,k}}} \right|^2}  \notag\\
 &  +\sum\limits_{i \ne k}^K {\left( {{{\left| {{\mathbf{h}}_k^H{{\mathbf{w}}_{c,i}} - x_{k,i}^c} \right|}^2} + {\lambda _{2,k}}{r_{k,{\text{th}}}}{{\left| {x_{k,i}^c} \right|}^2}} \right)} +{\lambda _{2,k}}{r_{k,{\text{th}}}}{\sigma ^2}.
\end{align}
To make dual function   $f_2\left( {{\lambda _{2,k}}} \right) = \mathop {\min }\limits_{x_{k,i}^c} {\cal L}_2\left( {x_{k,i}^c,{\lambda _{2,k}}} \right)$ bounded, we should make $1 - {\lambda _{2,k}} > 0$, i.e., ${\lambda _{2,k}} < 1$, since otherwise we can always set $x_{k,k}^c = \kappa {\bf{h}}_k^H{{\bf{w}}_{c,k}}$ and let $\kappa $ to be positive infinity, which will make   ${f_2}\left( {{\lambda _{2,k}}} \right)$ unbounded. This thus completes   the proof of Lemma~$2$.
\section*{Appendix C: \textsc{Proof of Theorem~2}}\label{appendix3}
Suppose that $\left\{ {{{{\mathbf{\bar W}}}_{c,k}},{{{\mathbf{\bar Z}}}_r}} \right\}$ are the converged solutions obtained by AO approach to problem  \eqref{P3}. We then construct another new solutions $\left\{ {{{{\mathbf{\hat W}}}_{c,k}},{{{\mathbf{\hat Z}}}_r}} \right\}$ that satisfy 
\begin{align}
&{{{\mathbf{\hat w}}}_{c,k}} = {\left( {{\mathbf{h}}_k^H{{{\mathbf{\bar W}}}_{c,k}}{{\mathbf{h}}_k}} \right)^{ - 1/2}}{{{\mathbf{\bar W}}}_{c,k}}{{\mathbf{h}}_k}, ~~{{{\mathbf{\hat W}}}_{c,k}} = {{{\mathbf{\hat w}}}_{c,k}}{\mathbf{\hat w}}_{c,k}^H, k\in{\cal K},\label{appendix3_1}\\
&{{{\mathbf{\hat Z}}}_r} = \sum\limits_{k = 1}^K {{{{\mathbf{\bar W}}}_{c,k}}}  + {{{\mathbf{\bar Z}}}_r} - \sum\limits_{k = 1}^K {{{{\mathbf{\hat W}}}_{c,k}}},  k\in{\cal K}. \label{appendix3_2}
\end{align} 
To prove Theorem~2, we need to prove: 1) ${\text{rank}}\left( {{{{\mathbf{\hat W}}}_{c,k}}} \right) = 1,{{{\mathbf{\hat W}}}_{c,k}} \succeq {\mathbf{0}},{{{\mathbf{\hat Z}}}_r} \succeq {\mathbf{0}}$; 2) the   objective value  obtained  by $\left\{ {{{{\mathbf{\hat W}}}_{c,k}},{{{\mathbf{\hat Z}}}_r}} \right\}$ in  \eqref{P3_obj} remains unchanged ; 3) all constraints \eqref{P3_const1}-\eqref{P3_const3}  are still satisfied. 

First, based on \eqref{appendix3_1}, it is not difficult to check that the newly constructed solutions ${{{\mathbf{\hat W}}}_{c,k}}, k \in {\cal K}$ are rank-one and positive semidefinite, i.e., satisfy  ${\text{rank}}\left( {{{{\mathbf{\hat W}}}_{c,k}}} \right) = 1,{{{\mathbf{\hat W}}}_{c,k}} \succeq {\mathbf{0}}$. In addition, for any ${\bm \phi}  \in {{\mathbb C}^{{N_t} \times 1}} \ne {\mathbf{0}}$, we have 
\begin{align}
{{\bm \phi} ^H}\left( {{{{\mathbf{\bar W}}}_{c,k}} - {{{\mathbf{\hat W}}}_{c,k}}} \right){\bm \phi} & = {{\bm \phi} ^H}{{{\mathbf{\bar W}}}_{c,k}}{\bm \phi}  - {\left( {{\mathbf{h}}_k^H{{{\mathbf{\bar W}}}_{c,k}}{{\mathbf{h}}_k}} \right)^{ - 1}}{\left| {{{\bm \phi} ^H}{{{\mathbf{\bar W}}}_{c,k}}{{\mathbf{h}}_k}} \right|^2}\ge 0, \label{appendix3_4}
\end{align}
where the last inequality follows from identity  ${\left| {{{\bm \phi} ^H}{{{\mathbf{\bar W}}}_{c,k}}{{\mathbf{h}}_k}} \right|^2} \le \left( {{{\bm \phi} ^H}{{{\mathbf{\bar W}}}_{c,k}}{\bm \phi} } \right)\left( {{\mathbf{h}}_k^H{{{\mathbf{\bar W}}}_{c,k}}{{\mathbf{h}}_k}} \right)$ according to the Cauchy-Schwarz inequality. From \eqref{appendix3_4}, it indicates that  ${{{\mathbf{\bar W}}}_{c,k}} - {{{\mathbf{\hat W}}}_{c,k}} \succeq {\mathbf{0}}$. Thus, we can see from \eqref{appendix3_2} that ${{{{\mathbf{\hat Z}}}_r}}$ can be rewrote as the summation of  $K+1$ positive semidefinite matrices, it follows that ${{{\mathbf{\hat Z}}}_r} \succeq {\mathbf{0}}$.

Second,  the expression $\sum\limits_{k = 1}^K {{{{\mathbf{\hat W}}}_{c,k}}}  + {{{\mathbf{\hat Z}}}_r}$  can be recast as 
\begin{align}
\sum\limits_{k = 1}^K {{{{\mathbf{\hat W}}}_{c,k}}}  + {{{\mathbf{\hat Z}}}_r} = \sum\limits_{k = 1}^K {{{{\mathbf{\hat W}}}_{c,k}}}  + \sum\limits_{k = 1}^K {{{{\mathbf{\bar W}}}_{c,k}}}  + {{{\mathbf{\bar Z}}}_r} - \sum\limits_{k = 1}^K {{{{\mathbf{\hat W}}}_{c,k}}} 
 = \sum\limits_{k = 1}^K {{{{\mathbf{\bar W}}}_{c,k}}}  + {{{\mathbf{\bar Z}}}_r},\label{appendix3_3}
\end{align}
where the first equality follows from  \eqref{appendix3_2}. Thus, we have $\sum\limits_{k = 1}^K {{\text{tr}}} \left( {{{{\mathbf{\hat W}}}_{c,k}}} \right) + {\text{tr}}\left( {{{{\mathbf{\hat Z}}}_r}} \right) = \sum\limits_{k = 1}^K {{\text{tr}}} \left( {{{{\mathbf{\bar W}}}_{c,k}}} \right) + {\text{tr}}\left( {{{{\mathbf{\bar Z}}}_r}} \right)$, which  shows that the objective value remains unchanged.

Third, substituting \eqref{appendix3_1} into ${\mathbf{h}}_k^H{{{\mathbf{\hat W}}}_{c,k}}{{\mathbf{h}}_k}$, we have 
\begin{align}
{\mathbf{h}}_k^H{{{\mathbf{\hat W}}}_{c,k}}{{\mathbf{h}}_k} = {\mathbf{h}}_k^H{{{\mathbf{\hat w}}}_{c,k}}{\mathbf{\hat w}}_{c,k}^H{{\mathbf{h}}_k} = {\mathbf{h}}_k^H{{{\mathbf{\bar W}}}_{c,k}}{\mathbf{h}}_k, k \in  {\cal K}. \label{appendix3_5}
\end{align}
Combining \eqref{appendix3_5} with \eqref{appendix3_3}, we can readily check that  constraints \eqref{P3_const1}-\eqref{P3_const3} are all satisfied.  Based on the above results, we complete the proof of Theorem~2.
\section*{Appendix D: \textsc{Proof of Theorem~4}}\label{appendix4}
By setting  ${{\bf{Z}}_r} = {\bf{0}}$ in  problem \eqref{P3} and denoting the newly formulated problem as  problem \eqref{P3}\textit{-new}, it is not difficult to see that any  feasible  solutions to problem \eqref{P3}\textit{-new} are also feasible to problem \eqref{P3}. 
Denoted by  $\left\{ {{{{\bf{\tilde W}}}_{c,k}},{{{\bf{\tilde Z}}}_r}} \right\}$  the  feasible  solutions to problem \eqref{P3}. We can 
 always construct another 
 solutions to problem \eqref{P3}\textit{-new}, denoted by  $\left\{ {{{{\bf{\mathord{\buildrel{\lower3pt\hbox{$\scriptscriptstyle\frown$}} 
						\over W} }}}_{c,k}}} \right\}$,  satisfying 
\begin{align}
{{{\bf{\mathord{\buildrel{\lower3pt\hbox{$\scriptscriptstyle\frown$}} 
					\over W} }}}_{c,k}} = {{{\bf{\tilde W}}}_{c,k}} + {\alpha _k}{{{\bf{\tilde Z}}}_r},~\sum\limits_{k = 1}^K {{\alpha _k}}  = 1, {\alpha _k}\ge 0,\forall k,
\end{align}					
so that $\sum\limits_{k = 1}^K {{{{\bf{\mathord{\buildrel{\lower3pt\hbox{$\scriptscriptstyle\frown$}} 
						\over W} }}}_{c,k}}}  = \sum\limits_{k = 1}^K {{{{\bf{\tilde W}}}_{c,k}}}  + {{{\bf{\tilde Z}}}_r}$ and ${\alpha _k}{\bf{h}}_k^H{{{\bf{\tilde Z}}}_r}{{\bf{h}}_k} \ge 0$, which indicates that 
any  feasible  solutions to   problem \eqref{P3} are also feasible to problem \eqref{P3}\textit{-new} while with the same objective value. Thus, problem  \eqref{P3} is equivalent to problem \eqref{P3}\textit{-new}. This thus completes the proof.

\bibliographystyle{IEEEtran}
\bibliography{Radcom}

\begin{thebibliography}{10}
\providecommand{\url}[1]{#1}
\csname url@samestyle\endcsname
\providecommand{\newblock}{\relax}
\providecommand{\bibinfo}[2]{#2}
\providecommand{\BIBentrySTDinterwordspacing}{\spaceskip=0pt\relax}
\providecommand{\BIBentryALTinterwordstretchfactor}{4}
\providecommand{\BIBentryALTinterwordspacing}{\spaceskip=\fontdimen2\font plus
\BIBentryALTinterwordstretchfactor\fontdimen3\font minus
  \fontdimen4\font\relax}
\providecommand{\BIBforeignlanguage}[2]{{%
\expandafter\ifx\csname l@#1\endcsname\relax
\typeout{** WARNING: IEEEtran.bst: No hyphenation pattern has been}%
\typeout{** loaded for the language `#1'. Using the pattern for}%
\typeout{** the default language instead.}%
\else
\language=\csname l@#1\endcsname
\fi
#2}}
\providecommand{\BIBdecl}{\relax}
\BIBdecl

\bibitem{Rappaport2013Millimeter}
T.~S. Rappaport \emph{et~al.}, ``Millimeter wave mobile communications for {5G}
  cellular: It will work!'' \emph{IEEE Access}, vol.~1, pp. 335--349, May 2013.

\bibitem{parker2017digital}
M.~Parker, \emph{Digital Signal Processing 101: Everything you need to know to
  get started}.\hskip 1em plus 0.5em minus 0.4em\relax Newnes, 2017.

\bibitem{zheng2019Radar}
L.~Zheng, M.~Lops, Y.~C. Eldar, and X.~Wang, ``Radar and communication
  coexistence: An overview: A review of recent methods,'' \emph{IEEE Signal
  Process. Mag.}, vol.~36, no.~5, pp. 85--99, Sept. 2019.

\bibitem{Saruthirathanaworakun2012Opportunistic}
R.~Saruthirathanaworakun, J.~M. Peha, and L.~M. Correia, ``Opportunistic
  sharing between rotating radar and cellular,'' \emph{IEEE J. Sel. Areas
  Commun.}, vol.~30, no.~10, pp. 1900--1910, Nov. 2012.

\bibitem{mu2018liu}
F.~Liu, C.~Masouros, A.~Li, H.~Sun, and L.~Hanzo, ``{MU}-{MIMO} communications
  with {MIMO} radar: From co-existence to joint transmission,'' \emph{IEEE
  Trans. Wireless Commun.}, vol.~17, no.~4, pp. 2755--2770, Apr. 2018.

\bibitem{Sodagari2012projection}
S.~Sodagari, A.~Khawar, T.~C. Clancy, and R.~McGwier, ``A projection based
  approach for radar and telecommunication systems coexistence,'' in \emph{2012
  IEEE GLOBECOM, Anaheim, California, USA}, pp. 5010--5014.

\bibitem{li2016optimum}
B.~Li, A.~P. Petropulu, and W.~Trappe, ``Optimum co-design for spectrum sharing
  between matrix completion based {MIMO} radars and a {MIMO} communication
  system,'' \emph{IEEE Trans. Signal Process.}, vol.~64, no.~17, pp.
  4562--4575, Sept. 2016.

\bibitem{li2016mimoconference}
B.~Li and A.~Petropulu, ``{MIMO} radar and communication spectrum sharing with
  clutter mitigation,'' in \emph{2016 RadarConf, Philadelphia, PA, USA}, pp.
  1--6.

\bibitem{Hassanien2016Signaling}
A.~Hassanien, M.~G. Amin, Y.~D. Zhang, and F.~Ahmad, ``Signaling strategies for
  dual-function radar communications: an overview,'' \emph{IEEE Aerosp.
  Electron. Syst. Mag.}, vol.~31, no.~10, pp. 36--45, Oct. 2016.

\bibitem{Sturm2011waveform}
C.~Sturm and W.~Wiesbeck, ``Waveform design and signal processing aspects for
  fusion of wireless communications and radar sensing,'' \emph{Proc. IEEE},
  vol.~99, no.~7, pp. 1236--1259, Jul. 2011.

\bibitem{Hassanien2016dual}
A.~Hassanien, M.~G. Amin, Y.~D. Zhang, and F.~Ahmad, ``Dual-function
  radar-communications: Information embedding using sidelobe control and
  waveform diversity,'' \emph{IEEE Tran. Signal Process.}, vol.~64, no.~8, pp.
  2168--2181, Apr. 2016.

\bibitem{hassanien2016phase}
------, ``Phase-modulation based dual-function radar-communications,''
  \emph{IET Radar, Sonar \& Navigation}, vol.~10, no.~8, pp. 1411--1421, Apr.
  2016.

\bibitem{csahin2021index}
A.~{\c{S}}ahin, S.~S.~M. Hoque, and C.-Y. Chen, ``Index modulation with
  circularly-shifted chirps for dual-function radar and communications,''
  \emph{IEEE Trans. Wireless Commun.}, 2021, early access, doi:
  10.1109/TWC.2021.3117063.

\bibitem{liu2018towards}
F.~Liu, L.~Zhou, C.~Masouros, A.~Li, W.~Luo, and A.~Petropulu, ``Toward
  dual-functional radar-communication systems: Optimal waveform design,''
  \emph{IEEE Trans. Signal Process.}, vol.~66, no.~16, pp. 4264--4279, Aug.
  2018.

\bibitem{liu2021dual}
R.~Liu, M.~Li, Q.~Liu, and A.~L. Swindlehurst, ``Dual-functional
  radar-communication waveform design: A symbol-level precoding approach,''
  \emph{IEEE J. Sel. Top. Sign. Proces.}, vol.~15, no.~6, pp. 1316--1331, Nov.
  2021.

\bibitem{joint2018liu}
X.~Liu, T.~Huang, N.~Shlezinger, Y.~Liu, J.~Zhou, and Y.~C. Eldar, ``Joint
  transmit beamforming for multiuser {MIMO} communications and {MIMO} radar,''
  \emph{IEEE Trans. Signal Process.}, vol.~68, pp. 3929--3944, Jun. 2020.

\bibitem{hua2021optimal}
\BIBentryALTinterwordspacing
H.~Hua, J.~Xu, and T.~X. Han, ``Optimal transmit beamforming for integrated
  sensing and communication,'' 2021. [Online]. Available:
  \url{https://arxiv.org/abs/2104.11871.}
\BIBentrySTDinterwordspacing

\bibitem{WU2020towards}
Q.~{Wu} and R.~{Zhang}, ``Towards smart and reconfigurable environment:
  Intelligent reflecting surface aided wireless network,'' \emph{IEEE Commun.
  Mag.}, vol.~58, no.~1, pp. 106--112, Jan. 2020.

\bibitem{wu2020intelligentarxiv}
Q.~Wu, S.~Zhang, B.~Zheng, C.~You, and R.~Zhang, ``Intelligent reflecting
  surface aided wireless communications: A tutorial,'' \emph{IEEE Trans.
  Commun.}, vol.~69, no.~5, pp. 3313--3351, May 2021.

\bibitem{liu2021Reconfigurable}
Y.~Liu, X.~Liu, X.~Mu, T.~Hou, J.~Xu, M.~Di~Renzo, and N.~Al-Dhahir,
  ``Reconfigurable intelligent surfaces: Principles and opportunities,''
  \emph{IEEE Commun.Surveys Tuts.}, vol.~23, no.~3, pp. 1546--1577, 3rd quart.
  2021.

\bibitem{wu2019intelligentxx}
Q.~Wu and R.~Zhang, ``Intelligent reflecting surface enhanced wireless network
  via joint active and passive beamforming,'' \emph{IEEE Trans. Wireless
  Commun.}, vol.~18, no.~11, pp. 5394--5409, Nov. 2019.

\bibitem{pan2020multicell}
C.~{Pan}, H.~{Ren}, K.~{Wang}, W.~{Xu}, M.~{Elkashlan}, A.~{Nallanathan}, and
  L.~{Hanzo}, ``Multicell {MIMO} communications relying on intelligent
  reflecting surfaces,'' \emph{IEEE Trans. Wireless Commun.}, vol.~19, no.~8,
  pp. 5218--5233, Aug. 2020.

\bibitem{hua2020intelligent}
M.~Hua, Q.~Wu, D.~W.~K. Ng, J.~Zhao, and L.~Yang, ``Intelligent reflecting
  surface-aided joint processing coordinated multipoint transmission,''
  \emph{IEEE Trans. Commun.}, vol.~69, no.~3, pp. 1650--1665, Mar. 2021.

\bibitem{zhou2020intelligent}
G.~Zhou, C.~Pan, H.~Ren, K.~Wang, and A.~Nallanathan, ``Intelligent reflecting
  surface aided multigroup multicast miso communication systems,'' \emph{IEEE
  Trans. Signal Process.}, vol.~68, pp. 3236--3251, Apr. 2020.

\bibitem{wu2021irs}
Q.~Wu, X.~Zhou, and R.~Schober, ``{IRS}-assisted wireless powered {NOMA}: Do we
  really need different phase shifts in {DL} and {UL}?'' \emph{IEEE Wireless
  Commun. Lett.}, vol.~10, no.~7, pp. 1493--1497, Jul. 2021.

\bibitem{mengjoint2021}
M.~Hua and Q.~Wu, ``Joint dynamic passive beamforming and resource allocation
  for {IRS}-aided full-duplex {WPCN},'' \emph{IEEE Trans. Wireless Commun.},
  2021, early access, doi: 10.1109/TWC.2021.3133491.

\bibitem{chen2021irsaided}
\BIBentryALTinterwordspacing
G.~Chen, Q.~Wu, W.~Chen, D.~W.~K. Ng, and L.~Hanzo, ``{IRS}-aided wireless
  powered {MEC} systems: {TDMA} or {NOMA} for computation offloading?'' 2021.
  [Online]. Available: \url{https://arxiv.org/abs/2108.06120.}
\BIBentrySTDinterwordspacing

\bibitem{hui2019Reflections}
\BIBentryALTinterwordspacing
H.~Long \emph{et~al.}, ``Reflections in the sky: Joint trajectory and passive
  beamforming design for secure {UAV} networks with reconfigurable intelligent
  surface,'' 2020. [Online]. Available: \url{https://arxiv.org/abs/2005.10559.}
\BIBentrySTDinterwordspacing

\bibitem{hua2021UAVSymbiotic}
M.~Hua, L.~Yang, Q.~Wu, C.~Pan, C.~Li, and A.~Lee~Swindlehurst,
  ``{UAV}-assisted intelligent reflecting surface symbiotic radio system,''
  \emph{IEEE Trans. Wireless Commun.}, vol.~20, no.~9, pp. 5769--5785, Sept.
  2021.

\bibitem{sixian2020robust}
S.~Li, B.~Duo, M.~Di~Renzo, M.~Tao, and X.~Yuan, ``Robust secure {UAV}
  communications with the aid of reconfigurable intelligent surfaces,''
  \emph{IEEE Trans. Wireless Commun.}, vol.~20, no.~10, pp. 6402--6417, Oct.
  2021.

\bibitem{mu2020Exploiting}
X.~Mu, Y.~Liu, L.~Guo, J.~Lin, and N.~Al-Dhahir, ``Exploiting intelligent
  reflecting surfaces in {NOMA} networks: Joint beamforming optimization,''
  \emph{IEEE Trans. Wireless Commun.}, vol.~19, no.~10, pp. 6884--6898, Oct.
  2020.

\bibitem{9365004}
------, ``Capacity and optimal resource allocation for {IRS}-assisted
  multi-user communication systems,'' \emph{IEEE Trans. Commun.}, vol.~69,
  no.~6, pp. 3771--3786, Jun. 2021.

\bibitem{9380234}
M.~Fu, Y.~Zhou, Y.~Shi, and K.~B. Letaief, ``Reconfigurable intelligent surface
  empowered downlink non-orthogonal multiple access,'' \emph{IEEE Trans.
  Commun.}, vol.~69, no.~6, pp. 3802--3817, Jun. 2021.

\bibitem{buzzi2021foundations}
\BIBentryALTinterwordspacing
S.~Buzzi, E.~Grossi, M.~Lops, and L.~Venturino, ``Foundations of {MIMO} radar
  detection aided by reconfigurable intelligent surfaces,'' 2021. [Online].
  Available: \url{https://arxiv.org/abs/2105.09250.}
\BIBentrySTDinterwordspacing

\bibitem{Lu2021Target}
W.~Lu \emph{et~al.}, ``Target detection in intelligent reflecting surface aided
  distributed {MIMO} radar systems,'' \emph{IEEE Sensors Lett.}, vol.~5, no.~3,
  pp. 1--4, Mar. 2021.

\bibitem{vaca2021radio}
\BIBentryALTinterwordspacing
C.~J. Vaca-Rubio \emph{et~al.}, ``Radio sensing with large intelligent surface
  for {6G},'' 2021. [Online]. Available:
  \url{https://arxiv.org/abs/2111.02783.}
\BIBentrySTDinterwordspacing

\bibitem{buzzi2021Radar}
S.~Buzzi, E.~Grossi, M.~Lops, and L.~Venturino, ``Radar target detection aided
  by reconfigurable intelligent surfaces,'' \emph{IEEE Signal Processing
  Lett.}, vol.~28, pp. 1315--1319, Jun. 2021.

\bibitem{9364358}
Z.-M. Jiang, M.~Rihan, P.~Zhang, L.~Huang, Q.~Deng, J.~Zhang, and E.~M.
  Mohamed, ``Intelligent reflecting surface aided dual-function radar and
  communication system,'' \emph{IEEE Systems J.}, 2021, early access, doi:
  10.1109/JSYST.2021.3057400.

\bibitem{song2021joint}
\BIBentryALTinterwordspacing
X.~Song, D.~Zhao, H.~Hua, T.~X. Han, X.~Yang, and J.~Xu, ``Joint transmit and
  reflective beamforming for {IRS}-assisted integrated sensing and
  communication,'' 2021. [Online]. Available:
  \url{https://arxiv.org/abs/2111.13511.}
\BIBentrySTDinterwordspacing

\bibitem{9416177}
X.~Wang, Z.~Fei, Z.~Zheng, and J.~Guo, ``Joint waveform design and passive
  beamforming for {RIS}-assisted dual-functional radar-communication system,''
  \emph{IEEE Trans. Veh. Technol.,}, vol.~70, no.~5, pp. 5131--5136, May 2021.

\bibitem{liu2021joint}
\BIBentryALTinterwordspacing
R.~Liu, M.~Li, Y.~Liu, Q.~Wu, and Q.~Liu, ``Joint transmit waveform and passive
  beamforming design for {RIS}-aided {DFRC} systems,'' 2021. [Online].
  Available: \url{https://arxiv.org/abs/2112.08861.}
\BIBentrySTDinterwordspacing

\bibitem{fishler2006spatial}
E.~Fishler, A.~Haimovich, R.~Blum, L.~Cimini, D.~Chizhik, and R.~Valenzuela,
  ``Spatial diversity in radars—models and detection performance,''
  \emph{IEEE Trans. Signal Process.}, vol.~54, no.~3, pp. 823--838, Mar. 2006.

\bibitem{cui2014MIMO}
G.~Cui, H.~Li, and M.~Rangaswamy, ``{MIMO} radar waveform design with constant
  modulus and similarity constraints,'' \emph{IEEE Trans. Signal Process.},
  vol.~62, no.~2, pp. 343--353, Jan. 2014.

\bibitem{cheng2018MIMO}
Z.~Cheng, Z.~He, B.~Liao, and M.~Fang, ``{MIMO} radar waveform design with
  {PAPR} and similarity constraints,'' \emph{IEEE Trans. Signal Process.},
  vol.~66, no.~4, pp. 968--981, Feb. 2018.

\bibitem{zheng2018joint}
L.~Zheng, M.~Lops, X.~Wang, and E.~Grossi, ``Joint design of overlaid
  communication systems and pulsed radars,'' \emph{IEEE Trans. Signal
  Process.}, vol.~66, no.~1, pp. 139--154, Jan. 2018.

\bibitem{probing2007Stoica}
P.~Stoica, J.~Li, and Y.~Xie, ``On probing signal design for {MIMO} radar,''
  \emph{IEEE Trans. Signal Process.}, vol.~55, no.~8, pp. 4151--4161, Aug.
  2007.

\bibitem{target2008xu}
L.~Xu, J.~Li, and P.~Stoica, ``Target detection and parameter estimation for
  {MIMO} radar systems,'' \emph{IEEE Trans. Aerosp. Electron. Syst.}, vol.~44,
  no.~3, pp. 927--939, Jul. 2008.

\bibitem{liu2006lu}
------, ``Radar imaging via adaptive {MIMO} techniques,'' in \emph{2006 14th
  EUSIPCO, Florence, Italy}, pp. 1--5.

\bibitem{boyd2004convex}
S.~Boyd and L.~Vandenberghe, \emph{Convex Optimization}.\hskip 1em plus 0.5em
  minus 0.4em\relax Cambridge University Press, 2004.

\bibitem{shi2016joint}
Q.~{Shi}, M.~{Hong}, X.~{Gao}, E.~{Song}, Y.~{Cai}, and W.~{Xu}, ``Joint
  source-relay design for full-duplex {MIMO} {AF} relay systems,'' \emph{IEEE
  Trans. Signal Process.}, vol.~64, no.~23, pp. 6118--6131, Dec. 2016.

\bibitem{zhang2017matrix}
X.-D. Zhang, \emph{Matrix analysis and applications}.\hskip 1em plus 0.5em
  minus 0.4em\relax Cambridge University Press, 2017.

\bibitem{sidiropoulos2006transmit}
N.~D. Sidiropoulos, T.~N. Davidson, and Z.-Q. Luo, ``Transmit beamforming for
  physical-layer multicasting,'' \emph{IEEE Trans. Signal Process.}, vol.~54,
  no.~6, pp. 2239--2251, Jun. 2006.

\end{thebibliography}
\end{document}